\documentclass[]{interact}

\usepackage{epstopdf}
\usepackage[caption=false]{subfig}

\usepackage[numbers,sort&compress]{natbib}
\bibpunct[, ]{[}{]}{,}{n}{,}{,}
\renewcommand\bibfont{\fontsize{10}{12}\selectfont}
\makeatletter
\def\NAT@def@citea{\def\@citea{\NAT@separator}}
\makeatother

\theoremstyle{plain}

\theoremstyle{definition}

\theoremstyle{remark}

\usepackage[utf8]{inputenc}
\usepackage{enumerate}
\usepackage{mathtools}
\usepackage{algorithm,algorithmic}
\usepackage{blindtext}
\usepackage{graphicx}
\graphicspath{ {./images/} }
\usepackage{epsfig}
\usepackage{sidecap}
\usepackage{hyperref,stackrel}
\usepackage[normalem]{ulem}
\usepackage{color}
\usepackage{enumerate}
\usepackage{bm}
\usepackage{amsthm,amsmath}
\usepackage{amsfonts}
\usepackage{amssymb,wasysym}
\usepackage{graphicx}
\usepackage{enumitem}
\usepackage{lmodern}
\usepackage{braket}
\usepackage[mathscr]{euscript}

\usepackage{multirow}
\usepackage [english]{babel}
\usepackage [autostyle, english = american]{csquotes}
\begin{document}

\articletype{Review}

\title{A Comprehensive Review of Quantum Random Number Generators: Concepts, Classification and the Origin of Randomness}

\author{
\name{Vaisakh Mannalath\textsuperscript{a}\thanks{ Vaisakh Mannalath. Email: vaisakhmannalath@gmail.com}, Sandeep Mishra\textsuperscript{a}\thanks{ Sandeep Mishra. Email: sandeep.mtec@gmail.com}, Anirban Pathak\textsuperscript{a}\thanks{ Anirban Pathak. Email: anirban.pathak@gmail.com}}
\affil{\textsuperscript{a}Jaypee Institute of Information Technology, A-10, Sector-62, Noida, 201309, India}
}

\maketitle

\begin{abstract}
Random numbers are central to cryptography and various other tasks. The intrinsic probabilistic nature of quantum mechanics has allowed us to construct a large number of quantum random number generators (QRNGs) that are distinct from the traditional true number generators. This article provides a review of the existing QRNGs with a focus on their various possible features  (e.g., device independence, semi-device independence) that are not achievable in the classical world. It also discusses the origin, applicability, and other facets of randomness. Specifically, the origin of randomness is explored from the perspective of a set of hierarchical axioms for quantum mechanics, implying that succeeding axioms can be regarded as a superstructure constructed on top of a structure built by the preceding axioms. The axioms considered are: (Q1) incompatibility and uncertainty; (Q2) contextuality; (Q3) entanglement; (Q4) nonlocality and (Q5) indistinguishability of identical particles. Relevant toy generalized probability theories (GPTs) are introduced, and it is shown that the origin of random numbers in different types of QRNGs known today are associated with different layers of nonclassical theories and all of them do not require all the features of quantum mechanics. Further, classification of the available QRNGs has been done and the technological challenges associated with each class are critically analyzed. Commercially available QRNGs are also compared.
\end{abstract}

\begin{keywords}
 Quantum Random Number Generators, Self-Testing Protocols
\end{keywords}

\section{Introduction}

The need for randomness is eternal; even in the epics, we see games based on specific devices' random outcomes. In fact, lottery and gambling of some form or other were also in existence in all the early civilizations. With time the relevance of random numbers has been widened, and new applications of random numbers have been found in weather prediction to Monte Carlo simulation \cite{Metropolis1949,Karp1991,10.1145/234313.234327}, cryptography to statistical sampling \cite{6769090,1621063}. In daily life, knowingly or unknowingly, we often use the output of random number generators. For example, one-time passwords (OTPs) received for various purposes, pins provided by the bank, and CAPTCHAs that appear on your screen when you try to log in at some website are all expected to be the output of a random number generator. Naturally, there is intense interest among scientists and technologists in building random number generators, and various types of random number generators have been built and/or proposed in the recent past \cite{james1990review,herrero2017quantum,ma2016quantum,jacak2021quantum}. The output from the random number generator should ideally conform to the principles of uniformity and independence. i.e., the random numbers produced should be equiprobable and there should be no correlations between them.
Further, privacy is another crucial parameter for application in cryptography. Interestingly, many of these random number generators are not true random number generators (TRNGs) and can best be referred to as pseudo-random number generators (PRNGs). PRNGs \cite{james1990review,hormann2004automatic,hull1962random} are based on some deterministic algorithms which convert a small input string into a larger set of strings that satisfy specific tests such as NIST test \cite{rukhin2001statistical}, Die-Hard test \cite{marsaglia2008marsaglia}, etc. to be useful in certain situations. PRNGs can produce random numbers fast, but such random numbers are not truly random as if one knows the algorithm and the initial input string, then all the random numbers can be predicted. Such random numbers can be helpful in situations where privacy is not essential such as simulations and weather forecasting, but not so for the case of secure communication and computation. 

TRNGs are different from PRNGs in that it is based on an unpredictable physical process.The solution to get a true random number sequence is to exploit the randomness associated with certain natural phenomena such as atmospheric noise \cite{haahr1999introduction}, cosmic background radiation \cite{lee2017cosmic}, thermal noises \cite{zhun2001truly}, noises in electronic circuits \cite{hamburg2012analysis}, chaotic systems \cite{stojanovski2001chaos}, etc.  Such phenomena are so complex that they can produce non-deterministic random numbers. However, the quality of such random numbers is hard to quantify \cite{killmann2011proposal}. It is challenging to certify the source of randomness as we cannot discard the possibility that a mature theory in the future will be able to accurately model the physical phenomenon that appears too complex to model at the moment, and hence the generated random numbers become predictable. So, it is always good to rely on a source with intrinsic randomness associated with it. We all know that quantum mechanics is inherently non-deterministic \cite{bera2017randomness,yuan2015}, so a quantum system can serve as an ideal source to generate truly random numbers \cite{ananthaswamy2019turn}. Further, the laws of quantum mechanics can be used to quantify the quality of random numbers. The current quantum technology is quite mature that it is now possible to generate high-quality random numbers which can be used in unconditionally secure cryptography applications \cite{jacak2021quantum}. The random number generators which use quantum sources to produce a certifiable source of randomness are known as quantum random number generators (QRNGs) \cite{herrero2017quantum,ma2016quantum}. Quantum systems have fascinating facets such as the existence of superposition states, collapse on the measurement, entangled but space-like separated particles, the existence of non-local correlations, the existence of indistinguishable particles \cite{nielsen,pathak2013elements}. Such features are counter-intuitive to our classical minds. These features play pivotal role in making the quantum systems intrinsically random and, hence, provide the sources for a new class of random number generators (QRNGs). 

The early QRNGs mainly exploited the property of uncertainty principle of quantum mechanics to build random number generators based on the phenomena of  radioactivity \cite{doi:10.1063/1.1658698,ishida1956random,alkassar2005obtaining}, shot noise in electronic circuits \cite{Vincent_1970,Gude+1985+187+190,586025,847868}, etc. Such systems had some drawbacks. For example, we may mention difficulty associated with the handling of radioactive sources and differentiating between shot and thermal noises. Further, they were also slow. To overcome these shortcomings, scientists looked towards exploiting the features of quantum optics and built generators based single-photon detectors \cite{jennewein2000fast,doi:10.1063/1.2338830,Stefanov_2000,bronner2009demonstrating,grafe2014chip,nie2014practical,khanmohammadi2015monolithic,Yan2014,burri2014spads,doi:10.1063/1.4897485,furst2010high,jian2011two} as well as macroscopic photodetection  \cite{shen2010practical,Gabriel2010,PhysRevLett.59.278,symul2011real,Li_2011,Nie_2015,bustard2011quantum,bustard2013quantum}. Similarly, atomic systems were also used to demonstrate QRNGs \cite{Pironio2010,um2013experimental}, for example, by exploiting the spin noise \cite{PhysRevA.77.054101,PhysRevLett.95.216603,PhysRevLett.98.176401}.Provided a complete control over the setup, there even exists a QRNG proposal, offering exponential speed-up in public randomness testing. The protocol in $ref.$ \cite{jacak2020quantum} achieves this by generating a set of random sequences at once such that randomness of any one those sequences can be evaluated by testing just one sequence among the set. This reduces the complexity in testing the randomness of all the generated sequences. The devices mentioned till now work on the assumption of the trust on the vendors, but there may be situations in which one needs to certify whether the device is working properly or not. Further, the privacy of the output is of utmost importance for applications in unconditionally secure communications. So, the self-testing QRNGs are the next generational devices in which the user can test the devices for privacy and randomness of the output without assuming trust in the vendors. Such self-testing QRNGs usually exploit the nonlocal correlations present in quantum systems. Many interesting experiments on the development of such devices have already been reported in the last decade \cite{colbeck2009quantum,Colbeck_2011,PhysRevA.87.012335,Pironio_2013,giustina2013bell,brandao2016realistic}. The major drawback of self-testing devices is that they have very low generation speeds in comparison to trusted devices. So, the solution lies in between with current focus on semi-self testing devices \cite{Cao_2016,Marangon_2017,PhysRevA.99.062326,Xu_2019,Avesani2018,Cao_2015,PhysRevA.94.060301,Tavakoli2021May,Lunghi_2015,li2011semi} which try to optimize the generation rates with the certification.

\begin{figure}
    \centering
    \includegraphics[width=0.95\textwidth]{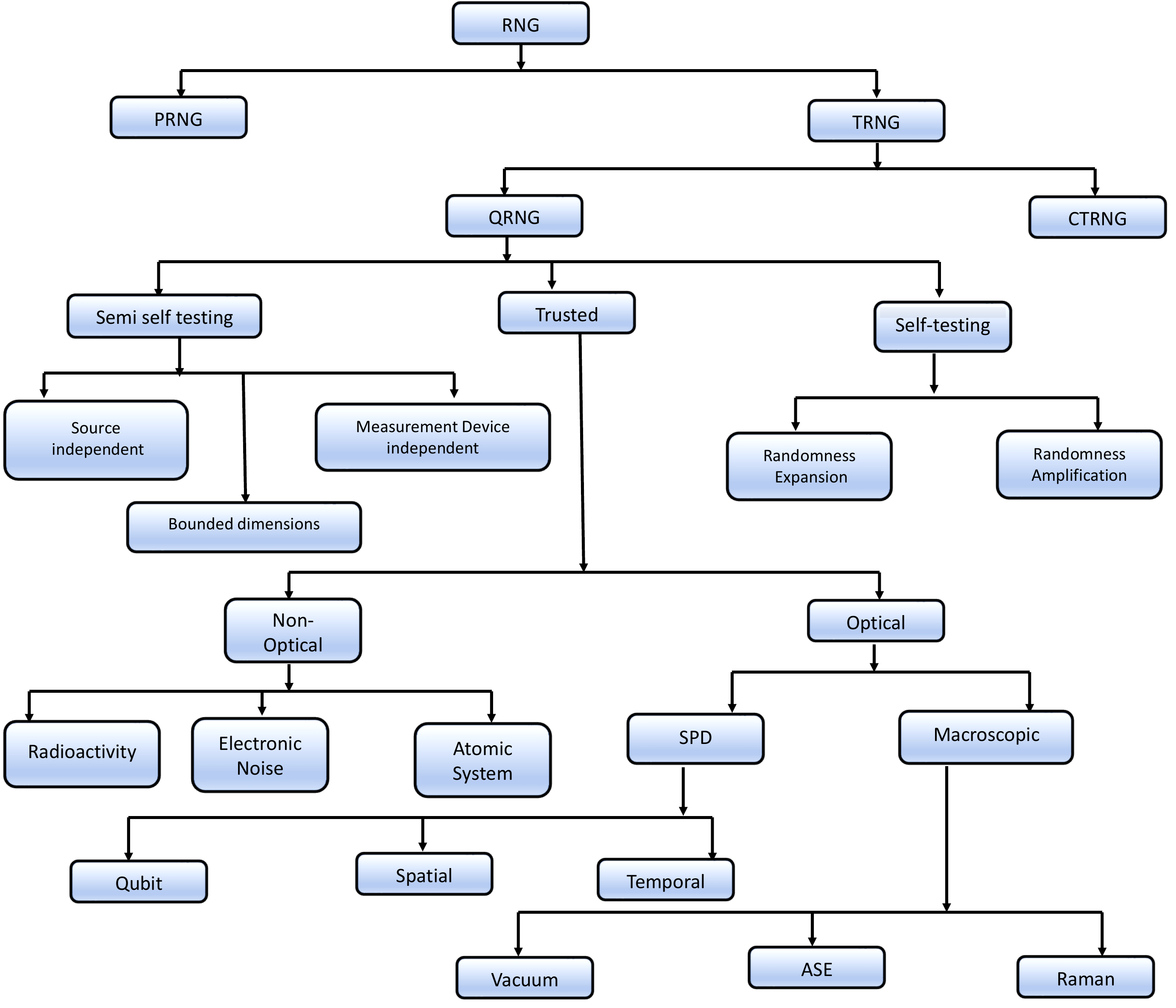}
    \caption{(Color online) Classification of random number generators. The classification is not strict and all-inclusive, with many generators overlapping in more than one class. RNG-Random Number Generator, PRNG-Pseudo Random number generator, TRNG-True Random Number Generator, QRNG-Quantum Random Number Generator, CTRNG-Classical True Random Number Generator, SPD-Single Photon Detector, ASE-Amplified Spontaneous Emission}
    \label{rng}
\end{figure}

Above discussion clearly indicates that there are different type of random number generators. A brief classification of known random number generators is shown in Figure \ref{rng}. Each class and subclass mentioned in Figure \ref{rng} will be discussed with appropriate importance in the subsequent sections. However, before we proceed to do that we would like to note that currently, the field of developing QRNG is very active with a slew of companies that have now come up with commercial QRNGs such as  ID Quantique \cite{QuantisQ20:online}, Toshiba \cite{httpswww1:online}, PicoQuant \cite{QuantumR34:online}, QuantumCTek \cite{QuantumR81:online} with excellent generation speeds and newer startups are coming up every year with new products. In short, exciting developments have been happening and this review aims to share that excitement with the readers. 

This review aims not only to track the progress in the development of QRNGs but also to explore the facets of randomness associated with available devices and their applications in specific systems and situations. In particular, we will explore the origin of the randomness of the currently available QRNGs as well as the futuristic devices from the perspective of a set of hierarchical axioms for quantum mechanics \cite{aravinda2019hierarchical} in which succeeding axioms can be regarded as superstructure built on top of the structure provided by the preceding axioms. The set of axioms to be used in this review can be briefly described as (Q1) incompatibility and uncertainty; (Q2) contextuality; (Q3) entanglement; (Q4) nonlocality and (Q5) indistinguishability of identical particles. Each of the mentioned axioms has an associated facet of the quantum systems. We know that quantum mechanics has different flavours of non-classicality, and these flavors have a multitude of different origins. The related debates among the researchers working in the foundations of quantum mechanics are still to be settled conclusively. Various people have devised mathematical constructs such as generalized probability theory (GPT) \cite{janotta2014generalized,barrett2007information}, generalized no-signalling theory (GNST) \cite{masanes2006general} and generalized nonlocal theory (GNLT) \cite{barrett2005nonlocal} to throw some light on the origin of quantum mechanical features. Our aim here is not to settle these hotly debated foundational questions, but to classify the QRNGs into new classes based on the origin of the randomness. We will come back to this in the latter part of the review after elaborating on the different classes of QRNGs shown in Figure \ref{rng}. {Further, another area of interest is regarding the certification of QRNGs with respect to the classical and quantum adversaries. Quantum adversaries are more powerful than their classical counterparts as they also have quantum resources in their hand. So, in this article we will also try to compare different QRNGs based on adversarial model.} 

This review is organized as follows: In Section \ref{sec:entropy}, randomness is mathematically defined in terms of entropy and it's shown that the randomness can be quantified. Subsequently, the differences between PRNGs and TRNGs are explicitly described in Section \ref{sec:pseudo&true}. Then, a  detailed overview of the trusted QRNGs is provided in Section \ref{sec:trusted}. Thereafter, we move towards the self-testing QRNGs in Section \ref{sec:selftesting} and semi-self testing QRNGs in Section \ref{sec:semi-selftesting}, respectively in sequence. Afterwards, we explore the origin of the randomness in QRNGs in Section \ref{sec:origin}. Finally, we conclude the review in Section \ref{sec:conclusions} with a brief mention of the commercially available QRNGs.

\section{Quantifying randomness via entropy\label{sec:entropy}}
Entropy measures the amount of randomness present in a system or equivalently the amount of information that can be gained from a system. A simple measure of randomness is given by the so-called Shannon entropy \cite{shannon1948mathematical}. It gives the average amount of information (measured in bits) that we can extract or gain from the system. For a random variable $X$ with a probability distribution $P_X(x)$, the Shannon entropy $H(X)$ is defined as
$$
H(X)=- \sum_{x\in X}P_X(x)\log _2P_X(x),
$$
where $X$ denotes the set of possible values of the random variable $x$. It measures the ignorance or information gained from knowing the value of the random variable. If some events are more likely than others, the average information gained from knowing an outcome will be less than a uniform distribution. A higher value of Shannon entropy implies that the probability distribution is relatively closer to uniform and vice versa. Clearly, we can see that if the random variable takes only a single value, the entropy is zero, while the entropy will be more if it can take many values with the entropy maximizing at the point when all possible outcomes are equiprobable. We can see an intrinsic connection between randomness and entropy of the system \cite{bera2017randomness,yuan2015,yuan2019}, i.e., the greater the entropy, the more randomness is. Consequently, entropy can be used as a measure of randomness. However, not all entropy measures are appropriate to use in the context of QRNGs. For example, Shannon entropy can only be used under the assumption that the random variables are independent and identically distributed in the asymptotic limit. This is hardly the case in practical setups where device imperfections come into play. For this reason, various other generalized entropies are also defined and connected to randomness \cite{e20110813}. 

A family of entropies that generalizes the Shannon entropy are the R\'enyi entropies \cite{renyi1961measures}, defined as
$$H_\alpha (X)=\frac{1}{1-\alpha}\log _2\sum_{x\in\mathcal{X}}P_X(x)^\alpha,$$
where $\alpha$ is the order of the entropy. R\'enyi entropy reduces to Shannon entropy in the limit $\alpha\rightarrow 1$. In general, for different orders of entropy $\alpha$ and $\beta$ satisfying $\alpha\leq\beta$  we have the relation between the R\'enyi entropies
$$
    H_\alpha(X)\geq H_\beta(X)
$$
for any distribution $X$. \\A widely used instance of it is the min-entropy $H_\infty(X)$ which is obtained in the limit $\alpha\rightarrow\infty$ as,
$$
H_\infty=-\log_2[
\max_{x\in\mathcal{X}}P_X(x)
].
$$
It is easy to see that min-entropy acts as a lower bound to all the R\'enyi entropies. Given a distribution $X$, $2^{-H_\infty(x)}$ gives the lower bound to the success probability of guessing one of the outcomes. In other words, min-entropy of $p$ implies that every event has a  bounded probability of $P_X(x)\leq2^{-p}$ to happen, and thus we can represent the probability distribution as distributions uniform in $p$ bits.    Since $k$ bits can be extracted from a distribution uniform in $k$ bits, for a distribution with min-entropy $p$, {for every outcome}, we can extract $p$ uniform bits \cite{4568168,zuckerman1990general}.

These measures of randomness via entropy can be extended to study joint distribution where the information content of a random variable may conditionally depend on other random variables. For example, a situation where part of the system is with $A$ and the rest with $B$. If we denote the combined system with the density matrix $\rho_{AB}$ in the Hilbert space $\mathscr{H}_{AB}=\mathscr{H}_A\bigotimes \mathscr{H}_B$, where $\mathscr{H}_A$ and $\mathscr{H}_B$ are the sub-spaces of the system accessible to $A$ and $B$, respectively, we can define the conditional min-entropy \cite{https://doi.org/10.3929/ethz-a-005115027} as
$$
    H_{\infty}(A \mid B)_{\rho}=\sup _{\sigma_{B}}\left(-\log _{2} \lambda\right),
$$
where $\sigma_{B}$ is the reduced density matrix of $B$ and $\lambda$ is the smallest real number such that
$$
\lambda I_{A} \otimes \sigma_{B}-\rho_{A B}
$$
is non-negative. 

Conditional min-entropy measures the information gained from the subsystem of $A$ given complete information about the subsystem of B. In other words, $2^{-H_\infty(A|B)_\rho}$ gives the lower bound to the probability of guessing outcomes of $A$ given complete knowledge of $B$ \cite{10.1109/TIT.2009.2025545}. If the distribution is uncorrelated, we get back the original min-entropy. Further, Konig and Renner \cite{5895072} introduced the concept of smooth min-entropy defined as
$$
H_{\infty}^{\epsilon}(A \mid B)_{\rho}=\sup _{\tilde{\rho}} H_{\infty}(A \mid B)_{\tilde{\rho}}
$$
with $\tilde{\rho}$ such that $\left\|\tilde{\rho}_{A B}-\rho_{A B}\right\| \leq \epsilon,\|A\|=\operatorname{tr} \sqrt{A^{\dagger} A}$. This entropy measures the valid outcomes for any particular sample rather than in the asymptotic limit. It becomes useful in the entropy estimation of randomness generators or in quantifying the capability of the randomness extractors to produce bits as close to a uniform distribution. However, in general, the estimation of min-entropy for a general unknown source is highly non-trivial \cite{watson2016complexity,Lyngso2002}. Repeated measurements can only provide a crude estimate. In most cases, bounds for the entropy are generated using analysis done on the physical source. QRNGs provide a clear advantage in this aspect. Quantum theory describes the randomness source precisely, compared to other generators based on physical processes, say, atmospheric noise. Even in the presence of classical noise or an eavesdropper, available min-entropy can be bounded using the accurate predictions provided by quantum theory. The existence of good randomness sources with easily quantifiable entropy is not enough if we don't have an efficient way for randomness extraction. Most random number generators with physical sources of entropy do not produce uniform bits; one needs to extract the available min-entropy into a uniform string using a randomness extractor for its practical use. In order to generate secure random numbers at high rates, various practical challenges have to be overcome, and various assumptions used during entropy estimation have to be satisfied. We will expand more on this in the subsequent sections.

\section{Pseudo and true random number generators\label{sec:pseudo&true}}
Deterministic algorithms can create strings of numbers that try to imitate the distribution statistics of random numbers. These algorithms constitute the so-called pseudo-random number generators (PRNGs). On the other hand, if the source of random numbers is unpredictable physical events, they are called true random number generators (TRNG). Albeit not being truly random, PRNGs have the advantage of being extremely fast, and for many applications, it is sufficient to have numbers generated by PRNG. In general, PRNGs use deterministic algorithms to convert a small string of input bits ($seed$)  to a larger string of bits that satisfies a specific set of randomness tests that a genuinely random sequence should satisfy \cite{l2012random}. Note that with knowledge of the initial seed and the algorithm used, one can easily predict the output of the PRNG \cite{pub2001140,barker2012recommendation}. This demands the input seed to be random in each run of the PRNG. The standard approach is to use sequences of uniform distribution from which other distributions can be generated using various transformations \cite{hormann2004automatic,knuth1997art}. Usually, PRNGs are based on the principles of number theory. For example, Lehmer introduced \cite{lehmer1951mathematical} linear congruential generators, which generates random numbers using the recursive formula
$$
    X_{n+1}=(aX_n+b) \hspace{2em}mod\hspace{0.3em} m, \hspace{2em}n\geq 0,
$$
where $X_i$ is the $i^{th}$ digit; which is multiplied by $a$ ($0\leq a<m$),  $b$ ($0\leq b<m$) is used as the increment, and $m>0$ is used as the modulus to obtain $X_{i+1}$.  Output numbers are sensitive on these input parameters and are also periodic, though the period is so large that they can be approximated to be random. Thus, the PRNGs are modeled to have considerable repetition periods to be still used as a random sequence. One of the most common pseudo-random number generators used in scientific software nowadays is the MT19937, based on Mersenne Twister (linear shift feedback registers \cite{klein2013linear}) \cite{matsumoto1998mersenne}, having a period of $2^{19937}-1$.
PRNGs also need to satisfy additional criteria to be used in cryptographic applications. They follow stricter conditions  than previously mentioned generators to increase the security of use in cryptographic protocols and are usually referred to as the cryptographically secure PRNGs (CSPRNG) \cite{herrero2017quantum}. Such generators should satisfy forward and backward security, meaning that no algorithm can use the knowledge of a part of the generated sequence to predict any preceding or succeeding bits with a probability greather than achieved by random guessing \cite{yao1982theory,10.1145/3335741.3335751}. A notable example of that is the Blum-Blum-Shub generators \cite{doi:10.1137/0215025} which are based on the bi-prime factorization problem. 
$$
    X_{i+1}=X_{i}^{2} \quad \bmod N
$$
where $N=pq$ where $p$ and $q$ are prime numbers. Breaking this generator is shown to be equivalent to the bi-prime factorization problem, which is considered to be computationally secure \cite{10.1007/3-540-39568-7_17}.

\begin{figure}[tb]
    \includegraphics[width=\textwidth]{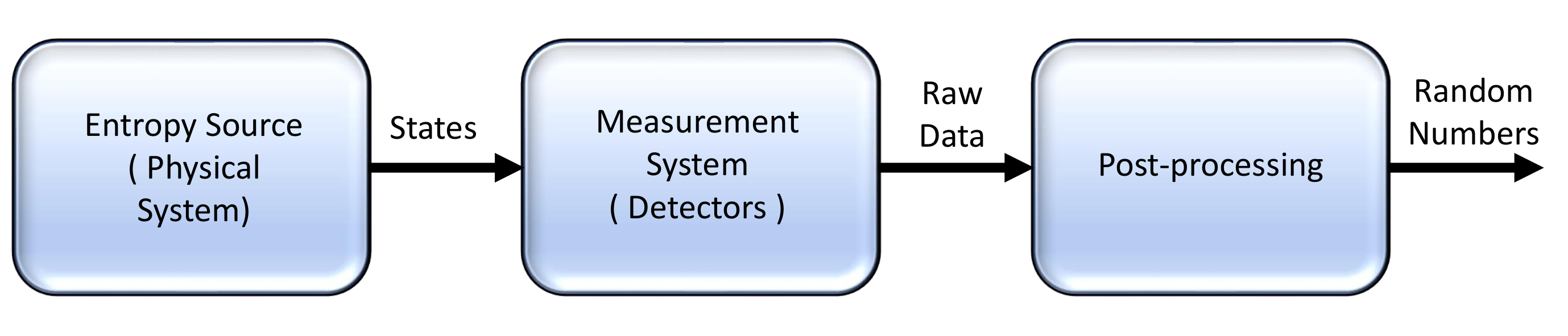}
    \caption{(Color online) Schematic of a true random number generator } 
    \label{fig1}
    \end{figure}

PRNGs provide an edge over other randomness generation methods for applications that require fast rates and reproducibility. Under appropriate conditions, they can be used for scientific simulations \cite{doi:10.1080/01621459.1949.10483310}, and cryptographic protocols \cite{ferguson2011cryptography} which requires large amounts of data that only need to mimic random statistics. Not all PRNGs are suitable for this task. It has been shown that linear congruential generators may have correlations undetected by standard randomness tests \cite{Marsaglia25}. Low-quality randomness has been shown to adversely affect the outcomes of scientific simulations \cite{10.1145/1276927.1276928}, and cryptographic protocols \cite{PhysRevA.86.062308}. Although PRNGs are faster than alternative random number generators, it is not entirely unpredictable. Since numbers generated using PRNGs can be replicated if one knows the initial value of the seeds, their security will be compromised if an adversary gets hold of it. To take care of this problem, one needs to look beyond the deterministic algorithms used by PRNGs. Certain physical processes are known to be good sources of randomness, whether due to our ignorance of the system or from something intrinsic. A TRNG uses such physical processes as its entropy sources and attempts to extract randomness. The schematic of a TRNG is shown in Figure \ref{fig1}.

 In fact, TRNGs are not without their own set of issues that hinder their wider adoption. They generally have a limited generation rate compared to PRNGs. They are limited by the physical processes that function as their randomness. The physical processes used in TRNGs could be anything from user's disk access times in an operating system, mouse motion, or keystrokes \cite{gutterman2006analysis}, to the noise in circuit electronics \cite{hamburg2012analysis}, and chaotic systems \cite{stojanovski2001chaos}. These processes provide a consistent entropy source and can also be complemented with other RNGs for added security \cite{becker2014stealthy}. The randomness they do generate is based on our ignorance of a particular physical system, and it is generally hard to quantify the amount of randomness generated. An example would be a TRNG using electrical noise in a circuit as its entropy source. Noise in such systems arises from the shot noise or the thermal noise. It is hard to differentiate between these two sources \cite{PhysRevB.47.16427} which becomes an issue when calculating the entropy. This failure to form tight bounds to the entropy makes testing the device for failures much more difficult. For the same reason, the failures occurring within a TRNG are harder to detect, making it prone to attacks \cite{killmann2011proposal}. QRNGs are a special case of TRNGs where the physical process that acts as a source of randomness is of quantum origin. For clarity, we will denote TRNGs that 
 can be clearly modelled with the quantum theory as QRNGs and the rest of the TRNGs as Classical TRNGs or CTRNGs.A clear distinction should be made between QRNGs and CTRNGs. The physical processes used in CTRNGs are often difficult to predict due to the incomplete knowledge of the system. Some so-believed CTRNGs, such as those using thermal noise and data in an electric circuit, may also take advantage of quantum phenomena. The only problem is that we do not know how to evaluate the amount of intrinsic randomness from basic physical principles. In this work, we only refer to QRNGs as devices whos working principles can be clearly modelled using quantum mechanics.  Since QRNGs are backed by a well-defined mathematical model of quantum theory one can get a certified source of randomness, which can be checked using the laws of quantum mechanics. The rest of the text will detail various QRNGs, including theoretical models, practical implementations, generation rates, etc.

\section{Trusted device QRNGs\label{sec:trusted}}
Trusted device QRNG assumes that the devices used to extract randomness and the provided source are well characterized. Their simplicity in theory and ease in implementation make them suitable for practical applications. They are only fit for application use if the implementation adequately fits the model. This is highly demanding in an experimental setup since controlling quantum systems is hard. These devices cannot certify whether the output bits are genuinely random or in the control of an adversary. This is why it is referred to as a trusted device. If you trust that your device is working as intended, you could choose appropriate quantum protocols to generate random numbers. In what follows, we will classify trusted device QRNGs into two subclasses: non-optical trusted device QRNGs and optical trusted device QRNGs. Under each sub-class, various possible implementations exist, and those are briefly described below with specific attention to their merits and demerits. To begin with, let us discuss different possible implementations of non-optical trusted QRNGs. 

\subsection{Non-optical trusted device QRNGs} 
This section will discuss various trusted device QRNGs that do not use optical elements in the randomness generation process. These include random number generators based on radioactive decays and other atomic systems and the systems based on electronic noise present in circuits.

\subsubsection{Radioactive Decay}
\label{radio}
\begin{figure}[tb]
    \includegraphics[width=\textwidth]{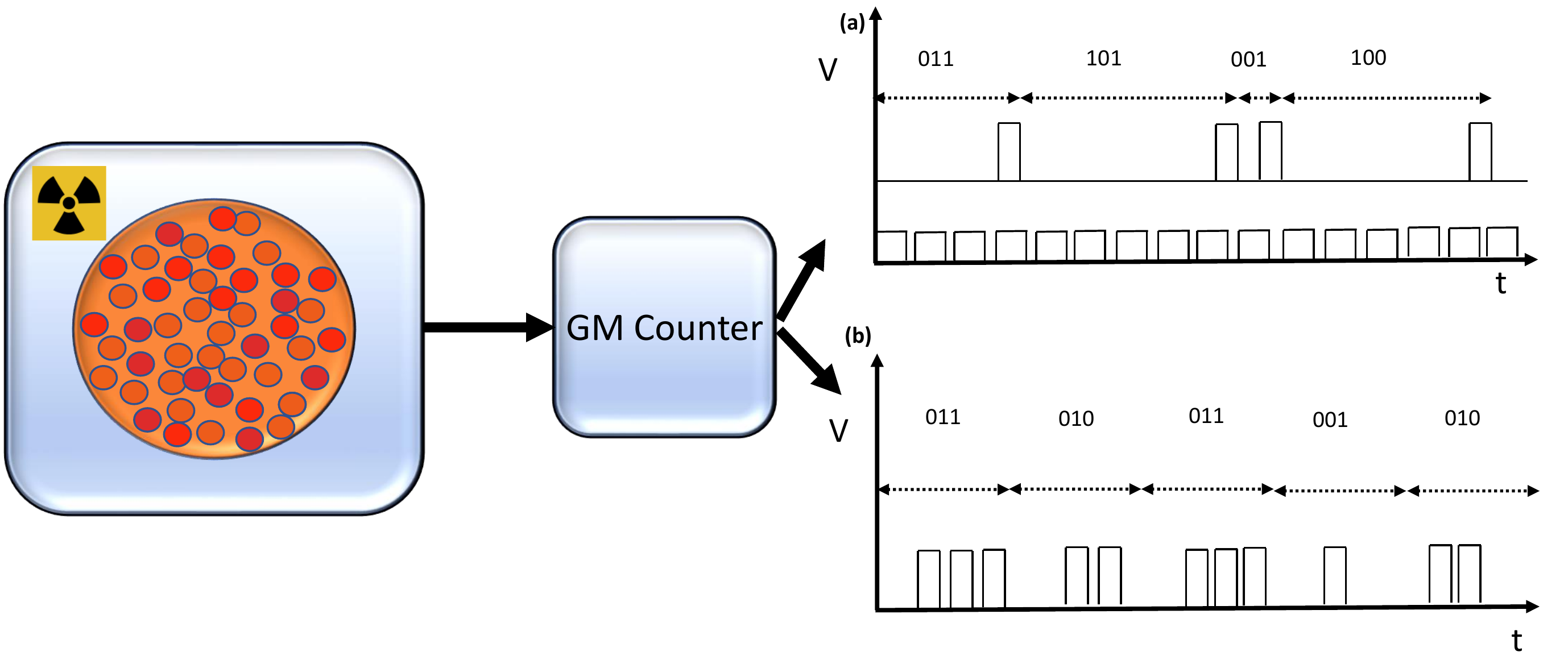}
    \caption{(Color online) Schematic of radioactivity-based TRNG with (a) Fast clock method: Random number generated equals the number of clock cycles between detections made by Geiger detector. \cite{herrero2017quantum}  (b) Slow clock method: The number of detection made by the Geiger detector in a fixed interval is taken to be the random number generated. \cite{herrero2017quantum}.} 
    \label{fig2}
    \end{figure}

 Radioactive decay of particles was one of the first quantum phenomena used to generate random numbers \cite{doi:10.1063/1.1658698}. Radioactivity can only be explained by using the uncertainty principle in quantum mechanics. Hence, QRNGs based on radioactivity effectively exploits the Q1 axiom only i.e. incompatibility and uncertainty.  The radioactivity based random number generators depended on sensitive Geiger-Müller (GM) tubes and well-characterized radioactive sources of $\alpha, \beta, \text { and } \gamma$ radiations. Detectors for $\beta$ radiations are much simpler and hence widely used. GM tubes produce a pulse for each detected particle using a Townsend avalanche \cite{1698087}. The probability of decay within the time interval `$dt$' is given by $$P(t)dt=\lambda_me^{\lambda_mt}dt,$$ where $\lambda_m$ is the decay constant of the particular radioactive source. The pulses form a Poisson distribution, and the exact rate depends on many factors like the half-life of the sample, position of the sample, state of gas in the GM tube. From the random arrival of pulses \cite{SILVERMAN1999265}, random numbers can be generated using a couple of different ways: $(i)$ Fast clock method \cite{Vincent_1970} (Fig. \ref{fig2}(a)), where the frequency of the clock is greater than the mean rate of detection and $(ii)$ Slow clock method \cite{doi:10.1063/1.1658698} (Fig. \ref{fig2}(b)), where the counts occur more frequently than a clock cycle. By measuring the clock cycles between two successive clicks, the randomness in the time of arrival is converted to bits. Every time detection is made, the fast clock reads and resets itself back to zero, and the corresponding time is used to generate a random number \cite{ishida1956random}. We can make the obtained distribution more uniform by taking the counts' parity (odd/even). If the counts are taken in binary, this is equivalent to taking the least significant digit or the coefficient of $2^0$ in the binary form. For example, if the number of counts is `2', it is written as `10' in binary, and the random number generated would be `0'. For an odd number, let us say `3' or `11' in binary; the generated number would be `1'. The slow clock method utilizes the number of counts in a fixed period to generate random numbers. One should restrict the number of counts to some integer $N$; a modulo counter. Furthermore, using modulo addition of different counts, one can obtain distributions with arbitrarily small bias \cite{doi:10.1063/1.1658698}. This will decrease the bias at the cost of reducing string length.

Some modern radioactive decay-based random number generators use semiconductor devices instead of GM tubes as it requires less voltage to operate and are convenient \cite{alkassar2005obtaining,duggirala2010radioisotope}. Even though the output signals are weaker, they can be amplified further. They simplify the design of the generators. Interestingly, the proposals in \cite{duggirala2010radioisotope} produce uniform distribution by using an RC circuit that converts an exponential or Poissonian random variable to one having uniform distribution.

QRNGs based on radioactive sources face severe drawbacks which limit their practical use. Since the randomness source is radioactive, it requires special care (improved safety measures) and knowledge. The radioactivity can also affect the detectors, decreasing their efficiency over time. Detector dead-time due to a buildup of ions inside the detector also poses a limitation. Due to the detector dead-time, GM tubes and semiconductor tubes need some time to recover their full detection abilities after a successful detection event \cite{1698087}. All of this needs to be taken into account while generating random bits, along with the appropriate post-processing steps \cite{Vincent_1971}. 

\subsubsection{Electronic Noise}
The noise present in electronic circuits can also be used as a source of entropy in extracting randomness. Typical random number generators based on electronic noises use circuit elements like resistors or diodes as the source of entropy. The noise typically arises due to the quantum nature of the charge carriers. Again, this is due to the uncertainty principle of quantum mechanics, hence only Q1 axiom is utilized. The noise, once generated, can be amplified and used for extraction. A simple method would be to compare the voltage output of the noisy element against a threshold value to generate random bits \cite{Vincent_1970,Gude+1985+187+190}. Another way would be to use the time of arrival methods detailed in Sec. \ref{radio}  \cite{586025,847868}. 

The noise present in such systems can be broadly classified as shot-noise \cite{https://doi.org/10.1002/andp.19183622304} and thermal noise \cite{PhysRev.32.110,PhysRev.32.97}. Shot noise arises due to the quantum nature of the current carriers and is hence truly probabilistic in nature \cite{REZNIKOV1998901}. It is caused
by carriers tunneling through quantum barriers created by $p-n$ junctions to create reverse or leakage current, which causes shot-noise. This jumping of quantum carriers shows up as voltage peaks across a  diode under adequately low current. Thermal noise, however, is caused by the motion of the carriers in response to the ambient temperature. Ideally, one would want to extract randomness from shot-noise since it's a  quantum process in the true sense. However, in practice, it is difficult to isolate these effects \cite{PhysRevB.47.16427} and hence extraction of randomness from shot noise suffers from this drawback. They are also known to have memory effects as voltage spikes depend on previous charge flow in the diode and thereby effects the quality of randomness extracted from it \cite{5967293}. Zener diodes and transistors have been known to predominantly exhibit shot-noise effects under the right conditions \cite{doi:10.1063/1.1809295, ReallyRe55:online}, and there are also commercial QRNGs that use shot-noise as a source of entropy \cite{Devicean84:online,Generato36:online}.  

\subsubsection{Atomic systems}
Apart from the approaches mentioned above for quantum random number generation, there have been proposals for random number generation using atomic systems. For example, in Refs. \cite{Pironio2010,um2013experimental},  trapped ions were used for random number generation. The experimental setups needed for random number generation using trapped ions are much more complex, and such setups produce random numbers with low generation rates. Still, various proposals for generating random numbers using atomic systems have been reported. Specifically, in Ref. \cite{PhysRevA.77.054101}, a QRNG based on the spin noise of an alkali metal vapor was proposed. The spin noise arises due to inherent quantum uncertainty and the noise arising due to the interaction between different atoms of the system.  Here again, only Q1 axiom is exploited to generate randomness. The signal acquired from such a process has been proved to be arising from a quantum noise and not from optical
pumping effects \cite{PhysRevA.75.042502}. In the proposed method, the spin noise of the Rubidium vapor was probed optically using a laser and converted to the polarization of a light beam. Under proper conditions, the spin noise can be separated from other background noises, generating random numbers. Further, increase in the bit generation rate have also been reported by the use of solid-state systems \cite{PhysRevLett.95.216603,PhysRevLett.98.176401}.

\subsection{Optical Trusted Device QRNG}
Optical QRNGs take advantage of the quantum nature of photons for generating random bits. Optics-based protocols are easier to implement than the methods mentioned above due to the ease of equipment availability and extensive research that has already been performed for various other purposes. Sources of entropy in this class of QRNGs are the light emitted from LASERs, LEDs, single-photon sources. The light is then manipulated using optical elements and eventually measured. Different protocols utilize different aspects of the quantum nature of light to extract randomness. We classify the different optical QRNGs based on the type of detector used as QRNGs using single-photon and QRNGs using macroscopic detectors.

\subsubsection{QRNGs using single photon detectors}
The devices belonging to this class utilize single-photon detectors (SPDs) for their functioning. Various techniques can be used to construct single-photon sources (often an approximate one), but we will not go into those details. For more details regarding single-photon generation and detection one may refer to  \cite{doi:10.1063/1.3610677,Buller_2009}. Instead of providing details of single-photon sources, in what follows, we will briefly describe a set of popularly used approaches for the realization of QRNGs using single-photon detectors.

\begin{figure}[tb]
    \includegraphics[width=\textwidth]{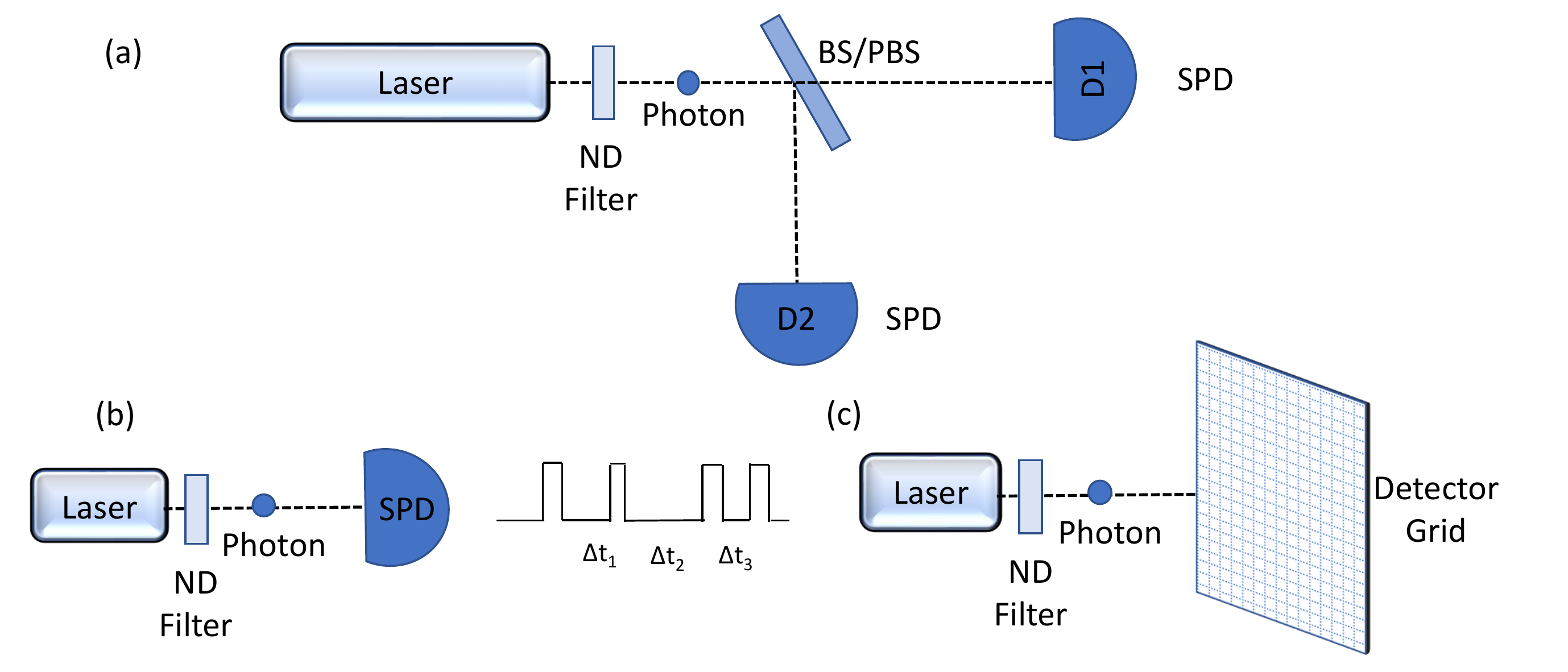}
    \caption{ (Color online) Schematic of (a) Qubit based QRNG where a photon in the $(\ket{H}+\ket{V})/\sqrt{2}$ state (Horizontal and  Vertical)  is  prepared. A polarizing beam splitter ($PBS$) selectively permits one of the vertical or horizontal states. In case of beam splitter ($BS$), a photon exists in the superposition state $(\ket{T}+\ket{R})/\sqrt{2}$ ( Reflected and Transmitted). A random bit can be generated by measuring the photon using two single-photon detectors (SPDs)  (b) Random bits generated by measuring the time interval between two detection events (c) QRNG based on measurements of photon spatial mode. The SPD array detects photons with some probability depending on the spatial position of the detected photon \cite{ma2016quantum}.} 
    \label{fig3}
    \end{figure}

\paragraph{Qubit state} These devices generate randomness by measuring qubits in superposition. The principle behind these devices lies in quantum theory's axioms that include collapse upon measurement. Hence, such devices can be explained based on the Q1 axiom only.  Specifically, the quantum state of a photon can be in a superposition of possible polarisation states or a superposition of possible paths. For example, if a photon is passed through a beam splitter (BS) then it exists in the superposition state of the sate in the reflected side ($\vert R \rangle$) as well as the state in the transmitted side ($\vert T \rangle$). Similar is the case when a linearly polarized photon (typically a photon polarized at an angle ${45}^{\rm{o}}$ with respect to the horizontal) inserts on a polarising Beam splitter (PBS)- a device that transmits horizontally polarized light and reflects vertically polarized light, then at the output the photon exists in a superposition of horizontal($\vert H \rangle$) and vertical($\vert V \rangle$) polarization (Fig \ref{fig3}(a)). Any quantum system can exist in a superposition of the basis states, and one can easily design a QRNG using this quantum property of photon along with the collapse on measurement postulate that states that a quantum state would collapse to one of the basis states upon measurement. In the computational basis $\{\vert 0 \rangle,\vert 1 \rangle \}$ with $\vert 0 \rangle$ and  $\vert 1 \rangle$ respectively, representing absence or presence of the photon, we can define
a state $|1\rangle_{1}|0\rangle_{2}$ which represents one photon in the first path and no photon in the second path, and a state $|0\rangle_{1}|1\rangle_{2}$ with the photon in the
second path and no photon in the first path. Then their superposition would look like,
$$
\frac{|1\rangle_{1}|0\rangle_{2}+|0\rangle_{1}|1\rangle_{2}}{\sqrt{2}}.
$$
Measuring the photons at both the arms would yield just one of the superposed states at random. Since there are only two possible outcomes, each detected photon would generate at most one random bit. There have been various implementations based on this principle involving beam splitters, and photo-multiplier tubes as detectors \cite{PhysRevLett.81.5039,jennewein2000fast}, Fresnel multiple prism  and avalanche photo-diodes \cite{doi:10.1063/1.2338830}, and the fiber-based implementations \cite{ManliXu2015}. The bit generation is limited to tens of Mbps \cite{jennewein2000fast}.

One can also think of a non-optical equivalent of these QRNGs. As a simple example, we may consider that we are using a cloud-based quantum computer of IBM \cite{IBMQ} which gives free access to superconductivity based 5-qubit quantum computer, where all the qubits are initially set to state $|0\rangle$, if we now place one Hadamard gate in each qubit line and subsequently perform measurement on each qubit line. Each measurement will randomly yield 0 or 1, and after each run of the experiment, we can have a 5-bit random number. IBM allows us to run an experiment 8192 times in a go and store the output in a sequence using the command  ``$result.get\_memory()$'' \cite{qiskit}. Thus, a single run of such a simple cloud-based experiment can give a $8192\times5=40960$ bit long string of random numbers. Of course, repeated runs or access to larger quantum computers can increase the size of the random numbers, thus generating the speed equivalently. This is noted just as an illustrative example to show that there are non-optical equivalents of the optical RNGs, and one can realize such a thing through cloud-based access of the quantum computer- a much more costly resource than a usual QRNG.

QRNGs based on the quantum state of photons suffer from practical drawbacks related to the detectors used. After every detection event, the detectors are inactive for a certain period during which they cannot detect photons. This period of inactivity is referred to as the dead-time of the detector. This will lead to correlations between the bits generated and increase the time required for acquisition. This could be avoided by using just one detector as in \cite{Stefanov_2000}. They connected two fibers of different lengths to the same detector and used the delay to distinguish between the two paths. Detectors could also detect multiple photons at once and may click even when no photons are present due to device imperfections. All of these effects influence the randomness and generation rate of the bits produced \cite{jennewein2000fast,rarity1994quantum,soubusta2003experimental,ManliXu2015,bronner2009demonstrating}. The random bit generation rates could also be improved if the
generator measures multiple paths \cite{grafe2014chip}. If the photons take more than two paths (say, $n$ paths), we can represent it as
$$
\left|W_{n}\right\rangle=\frac{|10 \cdots 00\rangle+|01 \cdots 00\rangle+\cdots+|00 \cdots 01\rangle}{\sqrt{n}}.
$$
Measurement in the path basis on the $n$ paths would result in one of the detectors clicking, yielding  $\log_2(n)$ bits of randomness. Unfortunately, this requires a much more complicated setup than the previously mentioned ones.

\paragraph{Temporal mode:} Temporal QRNGs generate randomness from the arrival time of photons. They are very similar to the QRNGs based on radioactive decay discussed in Section \ref{radio}, hence use only the Q1 axiom. Typically they have improved random number generation rate compared to radioactive decay since photon production is much faster. A standard temporal mode QRNG consists of a weak photon source, detectors, and timers. These devices generally use LEDs or LASERs for photon generation, and they follow an exponential distribution given by $\lambda e^{-\lambda t},$
where $\lambda$ is the average photon number
\cite{nie2014practical}. A simple approach to generate random numbers from the arrival time of photons would be to represent $t_1$
and $t_2$ as the arrival time of
consecutive pulses and assigning $1$ if $t_2 > t_1$ and $0$ if $t_1 > t_2$  (cf. Fig \ref{fig3}(b)),
giving a uniform random bit \cite{stipvcevic2007quantum,khanmohammadi2015monolithic}. The precision of the measured arrival time limits the randomness generation rate. Counts registered during a specified interval of time would be considered instantaneous due to the limited precision of the measuring instrument. The basic scheme that we mentioned would raise an error, and the counts would be unsuitable for random number generation. Yet another way of generation would be to count the number of clicks (cycles) using a fast (slow) clock, as in \ref{radio}. All such methods need to convert the randomness obtained from the exponential distribution to a more uniform bit sequence. There have also been implementations that use a uniform distribution of arriving photons. The generators in \cite{Wayne:10} uses special circuitry to convert the exponential distribution of incoming photons to a uniform one. The temporal-mode alleviates the impact of detection dead-time, unlike the methods based on superposition states (say, qubits) and generally has higher bit generation speeds. Generation rates are around around $109$ Mbps \cite{nie2014practical}  with an entropy of $5.5$ bits per detection.

\paragraph{Spatial mode:} Yet another method to generate random numbers is by measuring the spatial mode of a photon. \cite{Yan2014,burri2014spads}. This is usually done with a space-resolving detection system as shown in  (Fig \ref{fig3}(c)). Here also, only the Q1 axiom is sufficient for theoretical explanation. Experimental demonstration of the spatial mode technique using an array of detectors was done in \cite{doi:10.1063/1.4897485}. The spatial distribution of the light's intensity and the detector's efficiency affects the random numbers generated. Multiple random numbers can be generated per run with this method. Correlations formed by a large number of detectors make it impractical for commercial purposes. The practical implementation was able to generate up to $16$ bits per detection with an entropy of $0.999856$ at a speed of $8$ Mbps \cite{Yan2014}.
There also exist generators that use both time and
space uncertainties associated with the photons to generate randomness \cite{li2013true,thamrin2008photonic}.

\paragraph{Multiple photon-number states:} These devices generate randomness from measuring quantum states containing multiple
photons \cite{furst2010high,jian2011two,soares2014quantum,tisa2014high}. For example, by measuring the photon number of a coherent state,
$$\ket{\alpha}=e^{\frac{|\alpha|^2}{2}}\sum_{n=0}^\infty\frac{\alpha^n}{\sqrt{n!}}\ket{n}$$
which is  a superposition of photon-number states with $|\alpha|^2$ as the mean photon number, we can obtain random numbers that follows Poisson
distribution.
This is usually achieved by utilizing a multi-pixel detector arrangement. Since the photon number follows a Poisson
distribution, suitable post-processing is required to make the distribution uniform. Some generators use time
difference comparisons, similar to what is already discussed in the previous section \cite{ren2011quantum}, while
others generate randomness based 
 on the counted photon number, but the working of such devices can by explained by considering Q1 axiom only. A particular implementation of such a scheme of random number generation with $1.99$ random bits per detection and a speed of $144$ Mbps was reported in \cite{applegate2015efficient}.
The photon count variations have also been applied in random number generation using a mobile phone \cite{sanguinetti2014quantum}. Interestingly, these schemes are sensitive to photon number statistics and the detector's efficiencies.

The devices discussed so far use weak coherent sources as their single-photon source. The mean photon number of such sources should be attenuated around 0.1 to be considered a single-photon source, whereas a genuine single-photon source can achieve a mean photon number of 1. Solid-state single-photon emitters are a promising prospect in this direction, and much research has been focused on it over the past decade. Recently, there has been various implementations of QRNGs using solid-state single-photon emitters based on nitrogen vacancies in diamond \cite{chen2019single}, gallium nitride \cite{luo2020quantum}, hexagonal boron
nitride single-photon emitters \cite{white2020quantum}. Such devices have great potential in integrated quantum devices, although technical challenges pertaining to the complex mesoscopic environment of the solid-state need to be addressed \cite{aharonovich2016solid}.

\subsubsection{Macroscopic Photo-detection}
In this type of QRNGs, more classical quantities, like intensity, amplitude, etc., are measured instead of single photons to generate random numbers. Modeling these QRNGs is more complicated, and care must be taken to ensure that the dominant source of randomness is quantum. One upside is that these devices can circumvent the SPD dead-time limitation and hope to achieve a higher generation rate. The advantage obtained in this scenario is quite similar to the one obtained in the case of quantum key distribution protocols involving optical homodyne detection \cite{Grosshans2003} achieving higher key rates over a low loss channel. We will discuss two examples of this class based on vacuum noise and amplified spontaneous emission. 

\begin{figure}
    \centering
    \includegraphics[width=\textwidth]{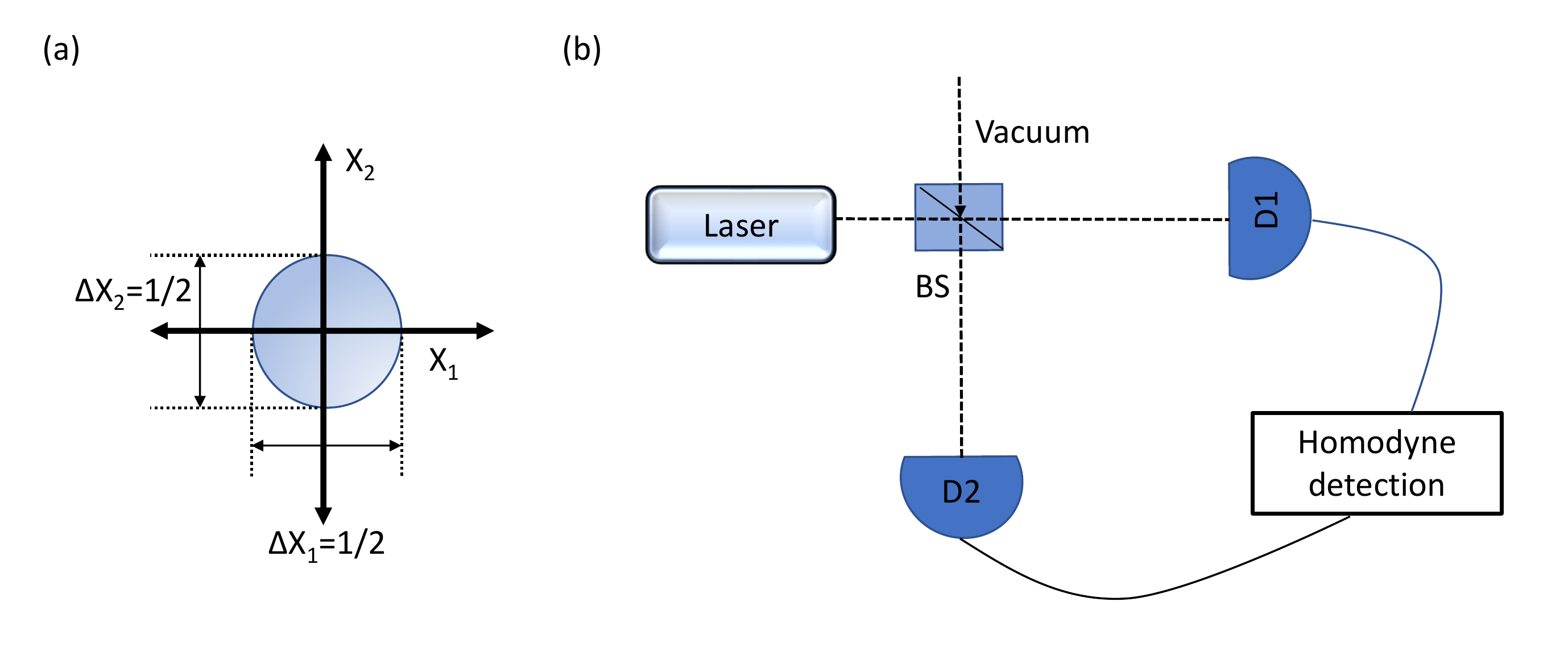}
    \caption{ (Color online) $(a)$ Phasor diagram for the vacuum state. The uncertainties in the two field quadrature are identical,
with $ \Delta X_1 =  \Delta X_2 = \frac{1}{2}.$  $(b)$ Schematic of QRNG based on vacuum noise measurements \cite{ma2016quantum}.}
    \label{vaccuum}
\end{figure}

\paragraph{Vacuum noise:} These QRNGs utilize the randomness in the zero-point fluctuation of the electromagnetic field, which is based on Q1 axiom. 
In quantum optics, the vacuum state is represented by a
two-dimensional Gaussian distribution centered at the origin with
an uncertainty of $1/4$ (see \autoref{vaccuum} (a) for its representation in the phase space). The amplitude and phase quadratures of the field can be measured repeatedly, thereby generating random numbers \cite{shen2010practical}. This uncertainty is present even though the vacuum state has zero photon number.
The vacuum state, the source of randomness, can be easily prepared with high fidelity. Being a
continuous variable, the measurement of the field quadrature of vacuum generates more than one random bit per measurement. In a particular example \cite{Gabriel2010}, $3.25$ bits
of random numbers were generated per measurement. A strong laser pulse is sent through one end of a symmetric beam splitter, and the other port is blocked to simulate a vacuum state. The output is then measured through two detectors, and the signal is obtained by taking the difference between the two outputs. The detection is done using the balanced homodyne detection scheme (cf. \autoref{vaccuum} (b)). Detector losses can be compensated by increasing the oscillator power. The random number generation rate of these devices is limited by the speed of the detector in the shot-noise region where the overall observed noise is dominated by vacuum noise \cite{PhysRevLett.59.278}. Realistically, shot noise is not the only noise present in the detectors, and an adversary can control the additional noises. Practical implementations have reached a generation rate of $3$ Gbps \cite{symul2011real}.\\

\paragraph{Amplified spontaneous emission:} Phase noise of amplified
spontaneous emission is quantum mechanical in nature and can be used to generate randomness \cite{Li_2011}. It is not limited by the shot noise region for detectors, as in the case of generators based on vacuum noise and is usually faster than them. In this particular class of QRNG, field quadratures of phase-randomized weak coherent states are measured to obtain random numbers \cite{Nie_2015}. Again, the randomness can be explained by considering the Q1 axiom. The uncertainty in the quadrature of the signal state is of the order $n\left\langle(\Delta \theta)^{2}\right\rangle$, where $n$ is the average photon number and $\left\langle(\Delta \theta)^{2}\right\rangle$ is the phase noise variance. If $n$ is sufficiently large, the phase uncertainty can be significantly larger than the vacuum noise. Hence, it is generally robust against detector noise. One way of implementation is using lasers, and heterodyne measurement \cite{Shakhovoy:20}. Interference of laser beams causes a variance in the output signal caused by the phase difference between the two input beams \cite{Qi:10}. Phase difference due to the unequal lengths of the interferometer is taken into account in the randomness generation process. Implementation involving pulsed laser sources where the phase difference between consecutive pulses is used to generate random numbers has also been demonstrated \cite{Jofre:11,Abellan:14}. 
\begin{figure}
    \centering
    \includegraphics[width=\textwidth]{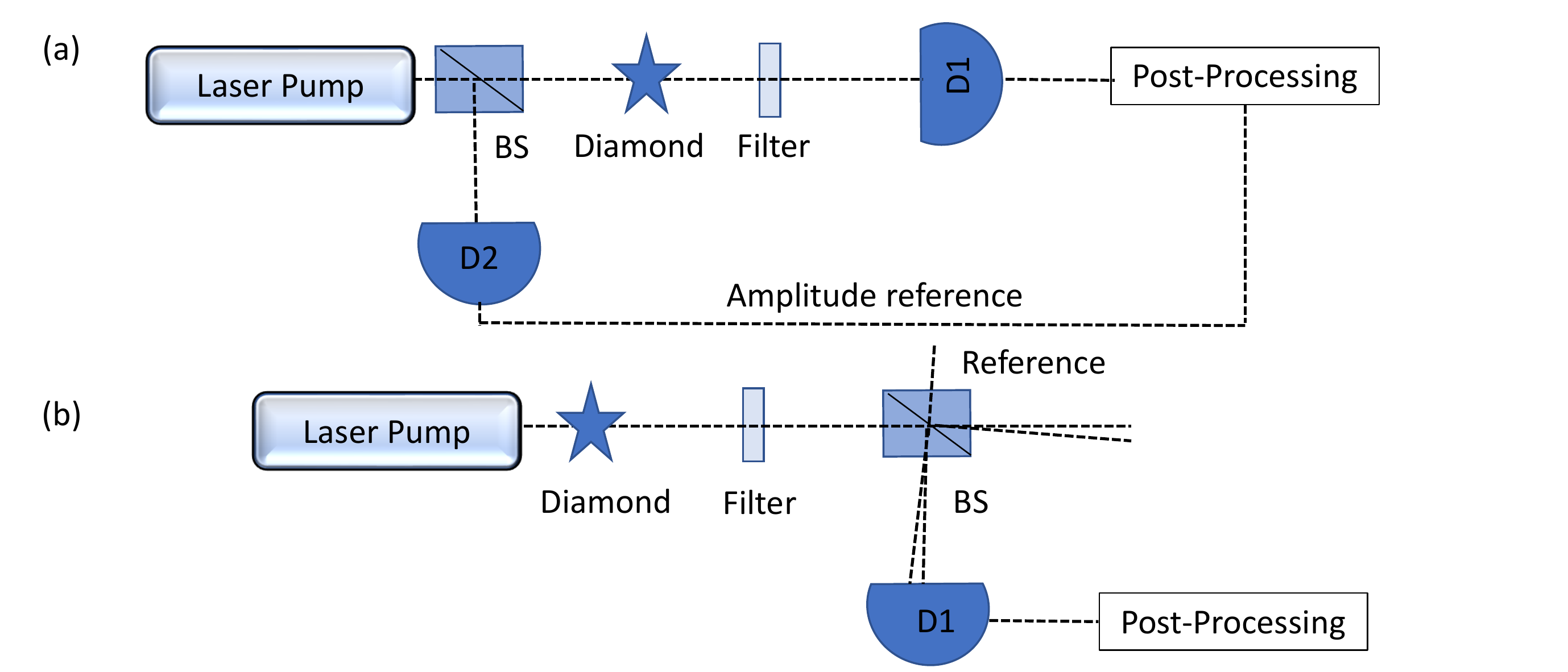}
    \caption{ (Color online) Schematic of QRNG using Raman scattering (a) with the amplitude fluctuations as a randomness source \cite{bustard2013quantum} (b) optical phase as randomness source \cite{bustard2011quantum}. }
    \label{fig:raman}
\end{figure}

\paragraph{Raman Scattering:} The inelastic scattering of photons by vibrational modes of molecules is known as Raman scattering. Various QRNGs use the phenomena of Raman scattering for gathering entropy in the form of randomized phase \cite{bustard2011quantum}, or amplitude \cite{bustard2013quantum} of the output field. We will be discussing random number generation based on two types of Raman scattering processes, namely spontaneous Raman scattering (SpRS) and stimulated Raman scattering (SRS) processes. In SpRS, a photon is scattered as it interacts with a molecular lattice that absorbs or creates a phonon to spontaneously generate photons of a higher frequency (anti-stokes) or lower frequency (stokes). In the case of SRS, some Stokes photons that have previously been generated by SpRS process amplify the emitted light as incoming photon interacts with the lattice. QRNGs based on Raman scattering work in a manner similar to ASE noise generators; SpRS photons that are produced at random from quantum noise are amplified in an SRS process \cite{PENZKOFER197955}. Fluctuations in the phonon field of the material induce the spontaneous emission to the Stokes
field \cite{RAYMER1990181} and the generated photons induce new Raman scattering processes, and the field is amplified to a macroscopic level. The quantum fluctuations at the initiating process can be viewed as uncertainty in amplitude (photon number)  and the optical phase at the output field. So, Q1 axiom is sufficient to explain the randomness os such devices. In a generator based on phase, difference \cite{bustard2011quantum}, optically pumped diamonds are used to generate Stokes field of random phase. These are then detected as interference patterns, the intensity patterns of which can be converted to a random sequence. For generators based on amplitude fluctuations \cite{bustard2013quantum} interference is not required, making it a simpler detection scheme \autoref{fig:raman}(a). The intensity values measured for the output field are compared with a reference and corrected. Intensity ranges are subsequently assigned to bit strings for generating random numbers. \\

Several other trusted device QRNGs exist, which do not fit into the abovementioned classes. In order to reduce the physical dimensions, QRNGs have been implemented in different substrates such as  indium phosphide (InP) \cite{abellan2016quantum}, lithium niobate (LN) \cite{haylock2019multiplexed}, silicon-on-insulator (SOI) \cite{prokhodtsov2020silicon}, silicon
nitride ($\text{Si}_3\text{N}_4$). Such devices have applications in photonic integrated circuits. These devices use electron-beam lithography to etch directional couplers that form the basis of the on-chip beam splitter. They are sensitive to temperature changes as the protocol depends on the substrate's refractive index, which is temperature-dependent. Sensitivity to the wavelength used and large optical losses are some of the other drawbacks. A brief history and comparison of several trusted device QRNGs implemented over the years are presented in Table \ref{trusttable}. They are compared based on the type of assumptions, usage of detectors, and their bit generation speeds.

\begin{table}[]
\begin{tabular}{|c|c|c|c|c|c}
\hline \text { Year } & \text { Physical principle } & \text { Detection } & \text { Speed } & \text { Adversary Model } 
\\\hline
2000 & \text { Electronic Noise}\cite{847868} & \text { IC } & 1.4 \text {Mbps } &\text{Classical Adversary} \\
 \text { 2000 } & \text { Spatial Superposition }\cite{jennewein2000fast} & \text { SPD } &1\text { Mbps }&\text{Classical Adversary}
\\
\text { 2000 } & \text { Spatial Superposition }\cite{stefanov2000optical} & \text { SPD } &100\text { Kbps }&\text{Classical Adversary}
\\
2005 & \text { Radioactive decay }\cite{alkassar2005obtaining} & \text { PIN photo diode } & 1.6\text {Kbps }&\text{Classical Adversary}  \\
\text { 2008 } & \text {  Time of arrival statistics }\cite{dynes2008high} & \text { SPD } &4\text { Mbps }&\text{Classical Adversary}
\\
2008 & \text{ Spin Noise}\cite{PhysRevA.77.054101} & \text { Polarimeter } & 1 \text{Kbps} &\text{Classical Adversary} \\
\text { 2009 } & \text {  Time of arrival statistics }\cite{wayne2009photon} & \text { SPD } &40\text { Mbps }&\text{Classical Adversary}
\\
\text { 2009 } & \text { Spatial Superposition }\cite{Kwon:09} & \text { PNRD } &500\text { Kbps }&\text{Classical Adversary}
\\
\text { 2010 } & \text { Photon number statistics }\cite{furst2010high} & \text { PNRD } &50\text { Mbps }&\text{Classical Adversary}
\\
\text { 2010 } & \text { Laser Phase Noise}\cite{Qi:10} & \text { Self-Heterodyne } &500\text { Mbps }&\text{Classical Adversary}
\\
\text { 2010 } & \text { Vacuum fluctuations}\cite{shen2010practical} & \text { Homodyne } &6.5\text { Mbps }&\text{Classical Adversary}
\\
\text { 2010 } & \text { Vacuum fluctuations}\cite{Gabriel2010} & \text { Homodyne } &12\text { Mbps }&\text{Classical Adversary}
\\
\text { 2010 } & \text {  Time of arrival statistics }\cite{Wayne:10} & \text { SPD } &110\text { Mbps }&\text{Classical Adversary}
\\
\text { 2011 } & \text { Vacuum fluctuations }\cite{symul2011real} & \text { Homodyne } &2\text { Gbps }&\text{Classical Adversary}
\\
2011 & \text { Electronic Noise }\cite{vartsky2011high} & \text { PIN photo diode } & 4 \text{Gbps}&\text{Classical Adversary}  \\
\text { 2011 } & \text { Raman Scattering }\cite{bustard2011quantum} & \text { Homodyne } &6\text { Kbps }&\text{Classical Adversary}
\\
\text { 2011 } & \text { Photon number statistics }\cite{ren2011quantum} & \text { PNRD } &2.4\text { Mbps }&\text{Classical Adversary}
\\
\text { 2011 } & \text { Time of arrival statistics }\cite{wahl2011ultrafast} & \text { SPD } &152\text { Mbps }&\text{Classical Adversary}
\\
\text { 2012 } & \text { Laser Phase Noise }\cite{Xu:12} & \text { Self-Heterodyne} &6\text { Gbps }&\text{Classical Adversary}
\\
\text { 2014 } & \text { Time of arrival statistics }\cite{nie2014practical} & \text { SPD } &96\text { Mbps }&\text{Classical Adversary}
\\
\text { 2014 } & \text {  Time of arrival statistics }\cite{doi:10.1063/1.2720728} & \text { SPD } &1\text { Mbps }&\text{Classical Adversary}
\\
\text {2014} & \text { Photon number statistics }\cite{tisa2014high} & \text { SPD } &200\text { Mbps }&\text{Classical Adversary}
\\
2014 & \text { Electronic Noise }\cite{wilber2013entropy} & \text { IC } & 32 \text{Mbps} &\text{Classical Adversary} \\
\text {2015} & \text { Laser Phase Noise }\cite{Nie_2015} & \text { Self-Heterodyne} &68\text { Gbps }&\text{Classical Adversary}
\\
\text { 2015 } & \text { Vacuum fluctuations }\cite{haw2015maximization} & \text { Homodyne} & 3.55\text { Gbps }&\text{Classical Adversary}
\\
2015 & \text{ Electronic Noise}\cite{ReallyRe55:online} & \text { PIN photo diode } & 23 \text{Kbps} &\text{Classical Adversary} \\
\text { 2015 } & \text { Photon number statistics }\cite{applegate2015efficient} & \text { PNRD} &143\text { Mbps }&\text{Classical Adversary}
\\
\text { 2015 } & \text {  Time of arrival statistics }\cite{khanmohammadi2015monolithic} & \text { SPD } &1\text { Mbps }&\text{Classical Adversary}
\\
\text {2016} & \text { Laser Phase Noise }\cite{Yang:16} & \text { Self-Heterodyne} &5.4\text { Gbps }&\text{Classical Adversary}
\\
\text {2017} & \text { Electronic Noise }\cite{Bernardo-Gavito2017Dec} & \text { SMU} &3.5\text { Kbps }&\text{Classical Adversary}
\\

\text { 2018 } & \text { Photon number statistics }\cite{Balygin2018} & \text { PNRD} &100\text { Mbps }&\text{Classical Adversary}
\\
\text { 2018 } & \text { Spatial Superposition }\cite{8329139} & \text { PNRD } &18.2\text { Mbps }&\text{Classical Adversary}
\\
\text { 2018 } & \text { Laser Phase Noise}\cite{2020} & \text { Self-Heterodyne } &600\text { Mbps }&\text{Classical Adversary}
\\
\text { 2018 } & \text { Vacuum fluctuations}\cite{Gehring2018Dec} & \text { Homodyne } &8\text { Gbps }&\text{Quantum Adversary}
\\
\text { 2019 } & \text { Vacuum fluctuations }\cite{https://doi.org/10.1002/que2.8} & \text { Homodyne} &1.5\text { Gbps }&\text{Classical Adversary}
\\
\text { 2019 } & \text { Vacuum fluctuations }\cite{doi:10.1063/1.5078547} & \text { Homodyne} &6\text { Gbps }&\text{Classical Adversary}
\\
\text { 2019 } & \text { Vacuum fluctuations }\cite{Guo:19} & \text { Homodyne} &8.25\text { Gbps }&\text{Classical Adversary}
\\
\text { 2021 } & \text {  Vacuum fluctuations }\cite{Gehring2021} & \text { Homodyne } &2.9\text { Gbps }&\text{Quantum Adversary}\\
\text { 2021 } & \text {  Electronic Noise }\cite{Aungskunsiri2021Aug} & \text { SMU } &90\text { Kbps }&\text{Classical Adversary}\\
\text { 2021 } & \text {  Time of arrival statistics }\cite{Zhang2021} & \text { SPD } &200\text { Kbps } &\text{Quantum Adversary}
\\
\hline
\end{tabular}
\caption{A comparison of various trusted QRNGs; SPD-single photon detector, PNRD-photon number resolving detector, IC-Integrated Circuit, SMU- Source Measure Unit} \label{trusttable}
\end{table}

\section{Self testing QRNGs\label{sec:selftesting}}
The methods described until now for the generation of random numbers rely on absolute trust in the used devices. However, this is not valid in realistic scenarios. Moreover, the trusted QRNGs exploit only the Q1 axiom for generation of randomness.   In practice, the devices may be defective or could even be in the control of an adversary and hence the generated random numbers become vulnerable. So, some form of testing system should be in place to verify the output sequence of the random number generator \cite{turan2018recommendation}. They can be modeled to monitor the internal state of the generator and notify the user of a sudden failure of the device or if its output is biased
\cite{fischer2012closer,bucci2005design}. However, now the user is required to trust the checking device, and we run into the same problem again. The implementations that we have detailed so far in the previous sections cannot rule out the possibility of a malicious vendor who supplies the checking devices. However, without an appropriate checking system, trusted device QRNGs cannot even differentiate the effects of classical noise from the generated numbers, thus putting a risk on the security of generated random numbers. Hence, it is better to include at least some self-testing systems in terms of security. We will now explore existing protocols in certifying generated random numbers from QRNGs.

If we assume that the vendor is not malicious but that devices may not be working perfectly, then the security of the protocols discussed so far can be increased by adding a checking system to them. For example, the 
quantum random number generator proposed in  \cite{saito2010randomness} uses a circuit that evaluates the time of arrival of pulses from a Geiger counter against a Poisson distribution for any discrepancies. Only sequences that pass this test are used in the extraction stage, while the others are discarded. \cite{fiorentino2006all} uses tomography techniques to estimate the matrix which describes the entropy source. Although this reduces the privacy of the numbers, it enables the user to estimate a lower bound on the min-entropy of the sequence based on the input quantum state. Further, the characterization of the input state offers protection against attackers that can alter the system's state. Vallone et al. \cite{PhysRevA.90.052327} provides a solution that does not require tomography of the system by randomly choosing between mutually unbiased bases for detection as they use the uncertainty principle to prove a bound on the amount of correlation the output sequence can have with the environment. The experimental demonstration was done using entangled photons, then measured based on polarization angles.

\begin{figure}
    \centering
    \includegraphics[width=\textwidth]{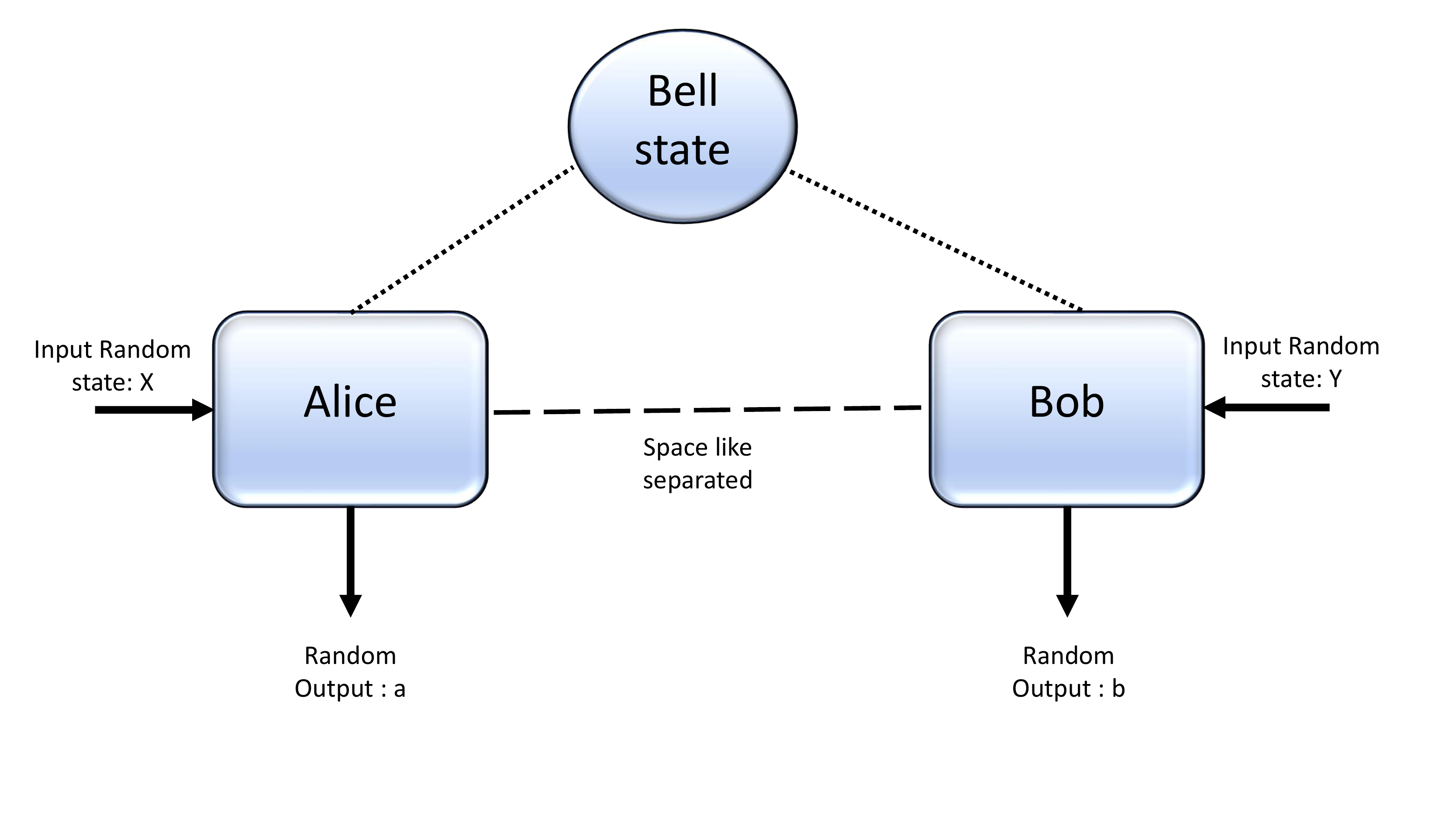}
    \caption{(Color online) Schematic representation of self-testing QRNG based on a Bell type inequality. Alice and Bob are space-like separated and share entangled state $\rho$ between them. For some random input random numbers $X$ and $Y$, Alice and Bob will respectively obtain the outputs $a$ and $b$. The device can be tested via the violation of some Bell-type inequality \cite{ma2016quantum}. }
    \label{bell}
\end{figure}

Another intriguing and conceptually interesting approach to certifying random numbers from a QRNG  is by ignoring the device's details and focusing only on the input-output statistics. Certification is done by comparing the output distribution against the distributions dictated by physical laws that one assumes to be true \cite{colbeck2009quantum}. Self-testing QRNGs exploit this idea in the form of witnessing non-locality \cite{bell1964einstein} by observing a Bell
inequality violation (\autoref{bell}). The term self-testing was first used in ref. \cite{Mayers1998Sep} in the context of quantum cryptography with imperfect sources.    Essentially,  such tests enables one to rule out local deterministic models of description for a particular setup by exploiting the Q4 axiom of quantum mechanics i.e. non-locality. Experimental results have supported quantum theory, and more sophisticated experiments are continually closing loopholes and alternative explanations. Classical randomness can be differentiated from genuine randomness from the amount of non-locality. Although this method has high credibility, it is hard to implement experimentally. It also suffers from the drawback of meager speed for generating random numbers.  Further, if we consider a completely device-independent setting, the devices should have randomized strategies and should invoke some form of non-locality. If the devices follow a deterministic protocol, then an adversary needs only to supply the devices with a set of data that passes the test that we have put in place. On the other hand, if the protocol is local, an adversary could supply the devices with the outputs with every choice of the input operation, rendering the test useless. Thus, it requires the user to access genuine random numbers and non-local states initially. In a sense, such protocols essentially expand the amount of randomness one has access to rather than generating randomness out of nothing, and thus it can be also be viewed as a device for random number expansion \cite{colbeck2009quantum}. 

As the discussion above indicates that there may be schemes for random number expansion, We will now look at various random number expansion protocols in detail. 

\subsection{Self testing Random number expansion} Random numbers generated using practical devices may depend on some classical variables, depending on the classical noise present in the system. This opens up security issues that have no easy fix.
Colbeck \cite{Colbeck_2011} was the first to suggest that untrusted devices can expand randomness. Since then, many protocols have been proposed, along with their security analysis. Self-testing randomness expansion protocols secure against
classical adversaries were proposed by Fehr et al. \cite{PhysRevA.87.012335}, and Pironio et al. \cite{Pironio_2013}, based on the Clauser–Horne–
Shimony–Holt (CHSH) inequality \cite{clauser1969proposed} which are based on exploiting the Q4 axiom. They used a setup with two devices. Each device had two possible settings, where choosing each setting yields one of the two outcomes. Based on the probability of these outcomes, one can form a correlation function. It admits a classical value that can be violated if one uses non-local (quantum) resources.
$$
    I=\sum_{x, y}(-1)^{x y}[P(a=b \mid x y)-P(a \neq b \mid x y)].
$$
This correlation function can be estimated by repeating the measurement a large number of times. It can then be used to estimate the min-entropy of the measurement output. Recently, Miller and Shi \cite{miller2015universal} demonstrated that as soon as the device violated CHSH inequality,  genuine randomness could be obtained. These protocols rely on the faithful realization of the Bell test. This is easier said than done, as the Bell tests are subject to locality and detection loopholes. Pironio et al. demonstrated
a proof-of-concept experiment of such randomness expansion protocol against classical adversaries \cite{Pironio_2010} in  an ion-trap system, which closes the detection loophole. However, they were only able to generate $42$ random bits over a period of one month, consuming more randomness than what was generated. Bit generation rates are typically low for these devices due to the increased complexity of the setup. Later proposals relaxed some conditions in order to have higher bit generation rates using optical components \cite{giustina2013bell}. Recently, the generation speeds have reached around the order of $10$ kbps, proven to be secure against quantum \cite{Liu2021,PhysRevLett.126.050503} and classical \cite{Shalm2021} adversaries.

In place of Bell inequality testing for non-locality, one can also consider other properties of quantum mechanics for verification. Generators reported in \cite{deng2013exploring,um2013experimental} used contextuality present in quantum mechanics  \cite{kochen1975problem} for generating random numbers.  It is related to the existence of non-commuting observables where the order of measurement is important, and there is no predefined model that can give the outcomes of two successive incompatible measurements. So, instead of exploiting the Q4 axiom which is at a higher level, the working of contextuality based devices can be explained by incorporation of Q2 axiom alone. These QRNGs work on untrusted devices but not on adversarial ones. Since contextuality tests do not require entanglement or space-like separation, their implementation is relatively easier than those based on Bell inequality. 
\\A fully device-independent protocol requires randomized settings as its input. A local-realist theory could explain the violation of Bell inequality if the measurement settings are correlated with the hidden variables. This is known as the Super-determinism loophole \cite{Larsson2014Oct}. The protocols mentioned so far require a perfect input random number to conduct loophole-free Bell inequality violations. In the following subsection, we will discuss how imperfect random numbers could generate perfect random numbers. In a sense, they `amplify' the randomness present in the input and are hence called randomness amplification protocol.

\subsection{ Self testing Randomness amplification} 
\label{randampDI}
A device-independent randomness expansion protocol requires an initial seed of randomness. Without it, the protocol would become deterministic, and it will be possible for an adversary to predict the outcome and change their strategy accordingly \cite{Colbeck_2012}. Randomness amplification protocol bypasses the need for a uniform seed. 
A model for weak randomness was described in ref. \cite{santha1986generating}, known as a Santha-Vazirani source. A sequence of random numbers $x_1x_2\cdots x_n$  of a  Santha-Vazirani source is called  $\epsilon$ free   if $$\epsilon \leq p\left(x_{j} \mid x_{1}, x_{2}, \ldots, x_{j-1}, e\right) \leq 1- \epsilon,$$ where $p\left(x\mid y\right)$ is the conditional probability and $e$ denotes all the classical variables that could possibly influence $x_{j}$. Given a weak source of random numbers, it is impossible to produce certifiable randomness classically \cite{santha1986generating}. Colbeck and Renner proposed the first randomness amplification protocol using quantum systems in \cite{Colbeck_2012}. They proved that any  weak
source of randomness with a critical threshold value of randomness could be amplified
into arbitrarily free random bits. Weak randomness sources are used to choose the measurement settings of quantum systems. Measurement outputs are then used to produce uniform random bits using chained Bell inequalities \cite{Braunstein1990Aug}. Later, Gallego et al. \cite{Gallego_2013} improved upon the result by showing that perfectly random bits could be generated from arbitrary weak randomness. Practical ways to use these protocols were developed in \cite{brandao2016realistic}, where only a limited number of independent devices were used. This is desirable since making sure that quantum devices are independent is difficult in practice. There have also been results on randomness amplification for more general sources taking any weak input with a bounded nonzero min-entropy\cite{chung2016general,PLESCH20142938,Bouda2014Sep}. Randomness amplification, apart from providing the initial seed in expansion protocols, also influence fundamental tests in science where initial randomness is required. The 'free will' loophole in Bell tests is an example of how imperfect randomness decreases the validity of such results \cite{koh2012effects}.

\section{Semi-self testing QRNGs\label{sec:semi-selftesting}} 
Realistically, parts of the random number generation device may be well characterized than others. In practice, one can choose to trust parts of the device and ignore details about other parts. This is especially relevant if we are only concerned about device imperfections, noisy channels, and not an adversarial situation. The security of such a device is intermediate to the trusted device and fully self-testing devices. It is a trade-off between practical trusted devices that provide high performance with low cost and low credibility and self-testing QRNGs with high credibility but low performance. One can consider a variety of these devices based on which part of the system is characterized.  We will now go over different models, namely source device-independent, measurement device-independent, and bounded dimension QRNGs.

\subsection{Source device independent QRNGs} As the name implies, in these devices, uncharacterized randomness sources are used. A trusted measurement device is used to monitor the source. Typically, measurement bases are chosen randomly from a complementing pair so that the presence of noise can be detected. This essentially means that uncertainty relations are exploited by such devices for which Q1 alone is sufficient. An initial random seed is required to choose the different measurement settings \cite{Zhang2021}. The generator could be modeled on a fixed set of assumptions or be made flexible, trading security for data rates \cite{Pivoluska2021Mar}. The source-independent QRNG of  \cite{Cao_2016} makes no assumptions about the dimension of the source (multiphoton emissions) and has demonstrated experimental bit rates up to $5\times10^3$bits/sec. Their generator takes into account the finite-key effect with the composable security definition. Composable security for a protocol is the most rigorous and functional security model since it can be used in conjunction with other protocols while remaining secure. Random number generators following a composable security definition can be used in arbitrary applications in cryptography, with no loss in security.   Recently, Drahi \emph{et. al.} have implemented a source-independent model generating composably secure quantum random numbers at a rate of $8.05$ Gbps. Certification measurements were performed on an untrusted light source mixed with a trusted vacuum. In the composable security definition they considered, protocols with certification failure $\epsilon_{1}$ and $\epsilon_{2}$ can be composed into a joint protocol with a total security parameter $\epsilon \leq \epsilon_{1}+\epsilon_{2}$. They were able to provide a composable security parameter of $10^{-10}$ for their generator. The generator of \cite{Marangon_2017} generates random numbers from the quadratures of an electromagnetic field without any assumption on the input state. They estimated a bound on the conditional min-entropy based on the entropic uncertainty principle for position and momentum observables of infinite-dimensional quantum systems. The randomness generation rates reach up to $1.7$ Gbit/sec. \cite{PhysRevA.99.062326} uses the phase randomized homodyne detection of gain-switched lasers \cite{PhysRevApplied.16.054012} to generate random bits at the rate of $270$ Mbits/sec. Higher rates were achieved in \cite{Xu_2019} with generation rates around $15$ Gbit/sec using quadrature fluctuations of a quantum optical field. They have quantitatively analyzed the effects of excess noise, finite sampling range, finite resolution, and asymmetric conjugate quadratures while developing the security analysis of the QRNG. Avesani et al. \cite{Avesani2018} experimentally demonstrated a source independent QRNG based on heterodyne detection with randomness generation rates over $17$ Gbps. Recently, a source-independent QRNG was also implemented using cloud quantum computers \cite{Li2021}.

\begin{figure}
    \centering
    \includegraphics[width=\textwidth]{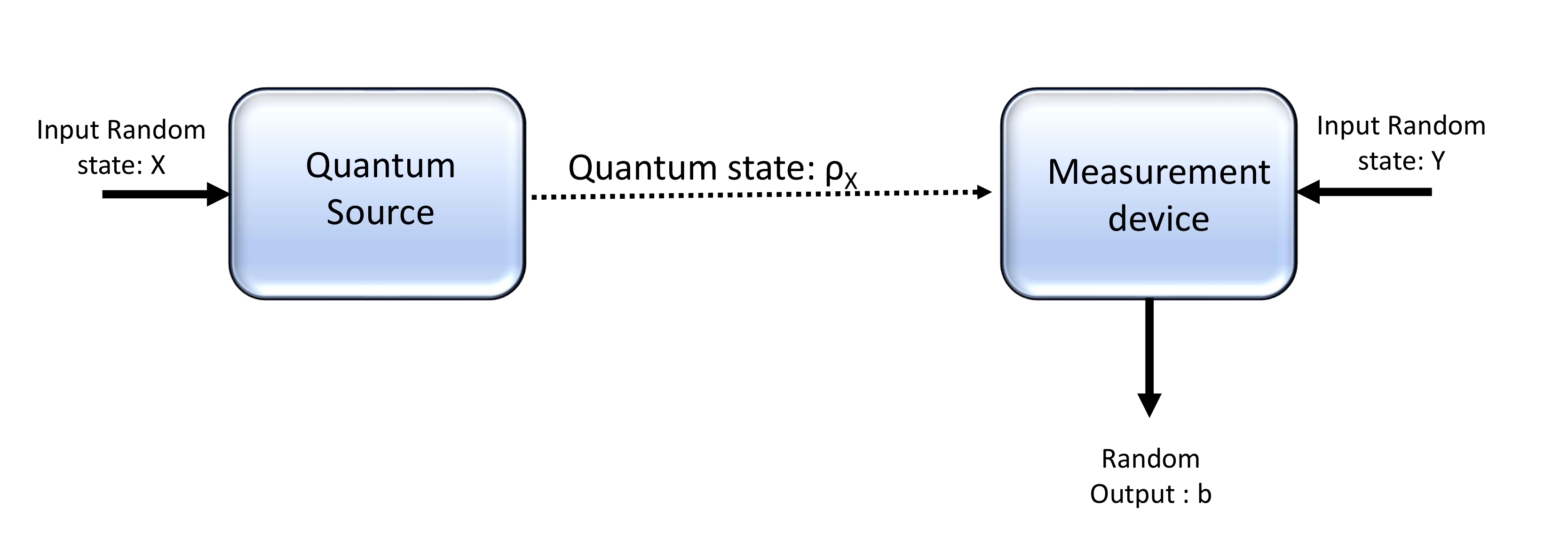}
    \caption{(Color online) Depending upon the input state $X$ emits a quantum state $\rho_x$. The measurement device measures the state $\rho_x$ based on the random input state $Y$ and gets an output $b$ \cite{ma2016quantum}}
    \label{semi}
\end{figure}

 \subsection{Measurement-device-independent QRNGs}
 If the input source is well characterized, whereas the measurement device is untrusted, we can still generate randomness using measurement device-independent QRNGs. In such devices, randomly chosen states are used to check the measurement device. Thus, in a sense, it is complementary to the source-independent QRNG. Again, here Q1 can explain the working of such devices. Cao et al. \cite{Cao_2015} proposed a protocol that is implementable using standard optical components. Nie et al. \cite{PhysRevA.94.060301} used an all-fiber setup to realize an MDI-QRNG based on time-bin encoding. Randomness certification in a network independent of the measurement devices used was considered in \cite{PhysRevA.95.042340}. \cite{Tavakoli2021May} recently introduced a prepare-measure framework based on bounded fidelity on the prepared states and subsequent certification of random numbers. The advantage of such QRNGs is that since the measurement devices are not required to be characterized, they remove possible attacks involving detector inefficiencies. The obvious downside is that any imperfection in the source will adversely affect the protocol since there is no way to account for that. As usual, high bit rates could be achieved using continuous-variable systems, but faces the challenges of continuous-variable entanglement witness and measurement
tomography.

\subsection{Self-testing QRNGs with bounded dimensions} Rather than trusting the source or the measurement device, one can assume some bound on the dimension of the system used and a witness to certify the randomness generation process. For example, the preparation and measurement devices could be thought of as operating in a two-dimensional quantum subspace \cite{Lunghi_2015}; the randomness of the protocol can be certified by adopting a particular dimension
witness. Any value of dimension witness greater than $0$ certifies the
randomness in the given example. Generation rates are typically low, reaching about $23$ bps. Generator of \cite{Lunghi_2015} assumes that the preparation and measurement device shares no classical randomness. They used a function presented in \cite{PhysRevLett.112.140407}, which acts as a `dimension witness'.
$$
\mathbf{W}=\left(\begin{array}{ll}
p(0,0)-p(1,0) & p(2,0)-p(3,0) \\
p(0,1)-p(1,1) & p(2,1)-p(3,1)
\end{array}\right)
$$
where $p(x, y)=p(b=0 \mid x, y)$. Whenever $\mathbf{W}>0$ randomness certification is proved to be possible.
Li. et al. \cite{li2011semi} proposed a protocol based on quantum random access codes (QRAC)\cite{ambainis2008quantum, PhysRevA.104.012420} which does not make any assumption regarding classical correlations between the preparation and measurement devices. They are based on the class of dimension witnesses introduced in \cite{PhysRevLett.105.230501}. The protocol in \cite{li2012semi} improves the randomness generation rates to its optimal value \cite{mannalath2021bounds} using a $3\rightarrow1$ QRAC. In a dimensionally bounded prepare-measure scenario, the randomness of the output can also be verified using non-inequality paradoxes such as Hardy’s paradox \cite{Hardy1992May} or Cabello’s paradox \cite{Cabello2002Feb}, as demonstrated in \cite{Li2015Aug}. A contextuality-based protocol was given in \cite{Pan2021Mar} using Mermin’s magic-square proof of Kochen–Specker contextuality \cite{Mermin1990Dec}. \\

Apart from the aforementioned protocols, there are various other semi-device independent random number expansion protocols based on different assumptions. A fair assumption to make in a practical prepare-measure setup would be a bound on the overlap between the prepared states \cite{VanHimbeeck2016Dec}; say a light pulse. A bound on the energy implies a lower bound on the overlap between the prepared states \cite{VanHimbeeck2019May}. If the states are not orthogonal, randomness can be extracted from the measurement outcomes of an unambiguous state discrimination protocol. Protocols based on this assumption include \cite{Tebyanian2021Sep,Tebyanian2021Dec,Rusca2020Jun,PhysRevA.100.062338}. Randomness certification from state discrimination was recently explored using non-contextuality as the notion of classicality \cite{Carceller2021Dec}. They have shown that quantum theory could certify more randomness than non-contextual ones. The notion of \emph{randomness amplification} explored in \autoref{randampDI} also have a semi-device independent analogue. Using just an additional assumption of either the dimension \cite{Zhou2015Aug} or energy bound \cite{Senno2021Aug} randomness amplification of weak sources have shown to be possible using quantum systems.

A brief history and comparison of several self-testing QRNGs implemented over the years are presented in Table \ref{DItable}. They are compared based on the type of assumptions, usage of the detectors, and their bit generation speeds.

\begin{table}[]
\begin{tabular}{|c|c|c|c|c|}
\hline \text { Year } & \text { Type } & \text { Detection } & \text { Speed} & \text{Adversary Model} \\
\hline 2010 & \text { Device Independent }\cite{Pironio_2010} & \text { lon-trap } & \text { Very Low } & \text{Quantum Adversary} \\
2013 & \text { Device Independent }\cite{giustina2013bell} & \text { SPD } & 0.4 \text { bps }& \text{Quantum Adversary}  \\
2015 & \text { Source Independent }\cite{PhysRevX.6.011020} & \text { SPD } & 5 \text{ Kbps} & \text{Classical Adversary} \\
2015 & \text{ Dimension Bound}\cite{Lunghi_2015} & \text { SPD } & 23 \text{ bps} & \text{Classical Adversary} \\
2016 & \text { Measurement Independent }\cite{PhysRevA.94.060301} & \text { SPD } & 5.7 \text{ Kbps} & \text{Classical Adversary} \\
2017&\text { Measurement Independent }\cite{PhysRevApplied.7.054018}&\text{SPD}& 16.5 \text{ Mbps}& \text{Classical Adversary}\\
2017&\text { Source Independent }\cite{PhysRevLett.118.060503}&\text{Heterodyne}& 1.7 \text{ Gbps}& \text{Classical Adversary}\\
2018&\text{Device Independent}\cite{BKG+18}&\text{SPD}&\text{ Very Low}& \text{Quantum Adversary}\\
2018&\text { Source Independent }\cite{Avesani2018}&\text{Homodyne}& 17 \text{ Gbps}& \text{Classical Adversary}\\
2018&\text{Device Independent}\cite{Liu2018}&\text{SPD}&181\text{ bps}& \text{Quantum Adversary}\\
2019&\text { Source Independent }\cite{PhysRevA.99.062326}&\text{Homodyne}& 270 \text{ Mbps}& \text{Classical Adversary}\\
2019&\text { Source Independent }\cite{Li2019}&\text{SPD}& 1 \text{ Mbps}& \text{Classical Adversary}\\
2019&\text { Source Independent }\cite{Michel2019Sep}&\text{Homodyne}& 8.2 \text{ Kbps}& \text{Classical Adversary}\\
2020&\text{Source Independent}\cite{PhysRevX.10.041048}&\text{Homodyne}&8.05\text{ Gbps} & \text{Classical Adversary}\\
2020&\text{Energy Bound}\cite{Rusca2020Jun}&\text{Homodyne}&145\text{ Mbps}& \text{Classical Adversary}\\
2020&\text{Device Independent}\cite{ZSB+20}&\text{SPD}&1.7\text{ bps}& \text{Quantum Adversary}\\
2021&\text{Device Independent}\cite{Liu2021}&\text{SPD}&13.5\text{ Kbps}& \text{Quantum Adversary}\\
2021&\text{Device Independent}\cite{Shalm2021}&\text{SPD}&3.6\text{ Kbps}& \text{Quantum Adversary}\\
2021&\text{Energy Bound}\cite{PhysRevApplied.15.034034}&\text{Heterodyne}&113\text{ Mbps}& \text{Classical Adversary}\\
\hline
\end{tabular}
\caption{A comparison of various self-testing QRNGs; SPD-single photon detector} \label{DItable}
\end{table}

\section{Origin of randomness in nonclassical theories\label{sec:origin}}
We have already discussed the limitations of PRNGs and CTRNGs in the previous sections. Further, we have delved upon the advantages offered by QRNGs and elaborated on the various class of QRNGs currently being available or in the development phase. This section will further explore the origin of randomness in the studied QRNGs. Precisely, we would aim to identify the nonclassical features of quantum mechanics responsible for the origin of randomness in a particular realization of QRNG and to investigate whether the kind of randomness obtained in that specific realization can also be obtained in a nonclassical toy theory which is not exactly quantum mechanics. A nonclassical toy theory $\mathcal{T}$ may be viewed as a theory having some features (axioms) which are not present in the classical world. For example, we may think of a nonclassical toy theory $\mathcal{T}_1$ that has all the features of the classical world, including "superposition," which is a manifestation of linearity of the theory and "collapse on measurement" postulate of quantum mechanics. As no other features (postulates) of quantum mechanics (e.g., entanglement, nonlocality) are included, $\mathcal{T}_1$ is a nonclassical theory (as a nonclassical feature, collapse on measurement is included) local (as nonlocality is not included) theory which is not equivalent to quantum mechanics (as all features of quantum mechanics is not present in  $\mathcal{T}_1$. Now it's easy to visualize that even in such a toy theory, the commercially available QRNG device called QUANTIS \cite{QuantisQ20:online} will work as its function involves only the collapse of a superposition state (qubit) on measurement using a reasonable basis. All the realizations of QRNGs are illustrated in Fig. \ref{fig3}(a)). $\mathcal{T}_1$ is just an example of nonclassical toy theory that can help us to lucidly describe the nonclassical origin of the unique features of QRNGs (more appropriately, TRNGs that are not CTRNGs or nonclassical RNGs) noted in this work. In what follows, we will follow this line of argument to profoundly investigate the origin of the so-called unique features of the QRNGs.

To begin with the analysis, we may note that quantum mechanics is intrinsically random and has various features such as entanglement, nonlocality, non-vanishing discord, etc., which are absent in the classical systems. Further, quantum mechanics has various interpretations, but the most accepted version is \textit{Copenhagen interpretation} \cite{stapp1972copenhagen}. In the Copenhagen interpretation, the physical states correspond to vectors in the Hilbert space, which evolve under unitary evolution. Even though Copenhagen interpretation has stood the test of time, many issues are still yet to be resolved, such as measurement problems. In contrast to other theories in physical sciences, quantum mechanics, as described by Copenhagen interpretation, is built upon the axioms that are more mathematical rather than having a physical motivation with terminology such as Hilbert space which cannot be directly observed. Various researchers are working on to derive (obtain) quantum mechanics from physically motivated axioms. Theories such as generalized probability theory (GPT) \cite{janotta2014generalized,barrett2007information}, generalized no-signalling theory (GNST) \cite{masanes2006general} and generalized nonlocal theory (GNLT) \cite{barrett2005nonlocal} have been developed in which classical mechanics and quantum mechanics come up as just one the cases. 

Generalized no-signaling theory (GNST) \cite{masanes2006general} is based on two axioms, namely (i) no signal can propagate faster than the speed of light, and (ii) existence of non-local correlations, which can violate Bell inequality. Quantum mechanics is a particular case of GNST, but there are other theories (other special cases of GNST) too, such as PR box \cite{popescu1994quantum}. Further, the PR box is known to be more non-local than quantum mechanics as the maximum value of Clauser-Horne-Shimony-Holt (CHSH) sum of correlations for QM is $2\sqrt{2}$ (Tsirelson bound) \cite{cirel1980quantum} while for that for PR box is 4. A natural question to consider is what bounds the non-local correlations of QM to $2\sqrt{2}$. Generalized non-local theory (GNLT) \cite{barrett2005nonlocal} tries to answer that by considering a set of theories that are only bound by no-signaling conditions and found that Tsirelson's bound restricts the use of quantum mechanics as a resource in distributed tasks. In another work, the principle of information causality is considered as the reason for separating quantum mechanics from the set of other non-signaling theories \cite{pawlowski2009information}. The principle of information causality states that using only local operations and classical communication (LOCC), the information accessible to one party is bounded by the information volume of communication from the other party. This principle separates the class of physical theories from non-physical theories. 

Generalized probability theory (GPT) \cite{janotta2014generalized,barrett2007information} is based on the framework in which the set of equivalent classes of preparation states and measurement outcomes form the basis. If $\omega$ represents an element of operationally equivalent class of preparation states and effect $e$ denotes  an operationally equivalent  class of one bit measurement outcome (`yes-no' type of answers), then $e(\omega)=p(e|\omega)$ denotes the probability that en experiment chosen from equivalence classes $e$ and $\omega$ produces a '1' (yes). Several authors have tried to describe quantum mechanics through a set of physically inspired axioms \cite{barrett2007information,hardy2001quantum}, but these axioms have no hierarchy. In a unique effort in this line of studies, some of the present authors explored the origin of quantum mechanics from the viewpoint of associating the axioms with different nonclassical features that have been experimentally observed \cite{aravinda2019hierarchical}. Further, they identified a hierarchy of five axioms where each succeeding axiom can be regarded as a superstructure built on top of the structure provided by the preceding axioms. These axioms have been named in as per the nonclassical features they are associated with: (Q1) incompatibility and uncertainty; (Q2) contextuality; (Q3) entanglement; (Q4) nonlocality and (Q5) indistinguishability of identical particles with each axiom associated with a relevant toy GPT model. It may be further noted that it's intrinsically assumed that any nonclassical theory $\mathcal{T}$  having one or more of the above axioms also contains classical features: (C1) Linearity, which leads to superposition in any linear theory (equivalently in any theory positing wave nature, e.g., classical optics); (C2) Tensor product space which in association with C1 gives classical entanglement (as pointed out by Simon et al. \cite{simon2010nonquantum}); (C3) Norm preserving evolution leading to unitarity which is satisfied by any theory with rotational invariance and satisfying C1 and C2; and (C4) No-signaling. In what follows, we would like to investigate which of these hierarchical axioms (or equivalently what kind of nonclassicality) are essential for realizing a specific type of QRNG and thus identifying the nonclassical toy theories that would allow the existence of such nonclassical RNGs.

 For the convenience of the discussion, we may describe five toy theories: $\mathcal{T}_{i}^{\prime}:i\in\{1,\cdots,5\}$, where a toy nonclassical theory  $\mathcal{T}_{i}^{\prime}$ would contain C1 to C4 and the hierarchical axioms up to Q\textit{i}. For example, $\mathcal{T}_{1}^{\prime}$ would be a nonclassical theory that will have features of classical world and Q1. A careful analysis of the various experimental methods that can be used to construct QRNGs would easily reveal that most of the QRNGs belonging to the class of trusted QRNG, $\mathcal{T}_{1}^{\prime}$ (i.e., the inclusion of Q1)  is sufficient. However, to obtain features like self-testing and semi-self-testing QRNGs, one would require theories placed hierarchically above as those would require the inclusion of a set of other axioms. Before we elaborate on that, we may note that the early QRNGs were mainly based on radioactive sources, and the random numbers generated from radioactive sources essentially follow the uncertainty principle. We all know that explanation of the radioactivity process requires the uncertainty principle. Thus, for a radioactivity-based nonclassical random number generator,  $\mathcal{T}_{1}^{\prime}$ would be sufficient. It may be noted that toy theory $\mathcal{T}_{1}^{\prime}$ is nonclassical, but it is a local theory, and such a theory would not only support radioactivity based RNGs it will also allow the construction of most of the trusted TRNGs that requires nonclassical features. As incompatibility implies collapse on measurement toy theory $\mathcal{T}_{1}$ described as an example above will be a particular case of $\mathcal{T}_{1}^{\prime}$ and QUANTIS and other nonclassical RNGs allowed in $\mathcal{T}_{1}$ will also be allowed in $\mathcal{T}_{1}^{\prime}$. Similar is the case for QRNGs based on the presence of shot noise in the electronic circuits, as it can be explained by quantum uncertainty relations alone. If one looks at QRNGs based on spin noise in atomic systems, one would also see that such systems would require Q1 only as spins noise arises due to inherent uncertainty relations. The QRNGs based on macroscopic photo-detection via vacuum noise exploit the uncertainty in the field quadratures. Similarly, only uncertainty relations are sufficient for randomness sources in QRNGs based on Raman scattering. In all the above scenarios, one can see that trusted QRNGs can be built in a nonclassical local theory $\mathcal{T}_{1}^{\prime}$. Another class of nonclassical local toy theory is $\mathcal{T}_{2}^{\prime}$, which has all features of $\mathcal{T}_{1}^{\prime}$ along with the contextuality Q2. Such a local nonclassical theory will be sufficient to support the self-testing devices which are based on contextuality  \cite{deng2013exploring,um2013experimental} by observing the violation of contextuality expressions \cite{kochen1975problem} for certification of the devices against the presence of any adversary. It is important to note that such certification is only local in nature. Now, let us look at the self-testing QRNGs. They exploit the features of entanglement and non-locality. We all know that any local realistic hidden variable theory cannot explain the quantum correlations. Self-testing devices work by focusing only on input-output statistics. So, the self-testing devices can be certified by the violation of Bell-type inequalities \cite{clauser1969proposed}. Consequently, a nonlocal toy theory $\mathcal{T}_{4}^{\prime}$ which includes Q1, Q2, Q3, and Q4 will be required for such devices. Note that even $\mathcal{T}_{1}^{\prime}$ is not equivalent to quantum mechanics as it does not require Q5. We conclude this section by noting that keys generated in quantum key distribution protocols are essentially random, and naturally, different protocols can be supported by different toy theories and analysis in the line of the present section in relation to the schemes of QKD can be found in our earlier work \cite{aravinda2019hierarchical,aravinda2017origin}.

\section{Conclusions\label{sec:conclusions}}

A comprehensive overview of the existing random number generation methods using quantum systems has been done in this article. We have described QRNGs of various kinds in the previous sections. The discussion above clearly establishes the advantages of QRNGs and the technological challenges associated with implementing different classes of QRNGs. It also establishes that despite some existing technological challenges, there are many alternatives to produce QRNGs. The associated technology is much more mature than the majority of other aspects of quantum technologies that are expected to have a substantial impact (say, a scalable quantum computer). Naturally, a large number of commercial products (QRNGs) have been launched. Specifically,  the development of QRNGs has advanced to a point where off-the-shelf QRNGs are now commercially available and not costly. A brief overview of the available products is provided in \autoref{tab:comtab}. The table lists commercially available QRNGs along with their bit generation speed, types of connectivity available, and the different tests passed by the corresponding QRNGs \cite{jacak2021quantum}. It is to be mentioned that the commercial QRNGs available now belong to the class of trusted devices. The next-generation devices belonging to the self-testing and semi-self-testing classes are still in the developmental stage, and their commercial utilization is expected sooner than later. 

\begin{table}
\resizebox{\columnwidth}{!}{\begin{tabular}{|c|c|c|c|c|c|}
\hline Company& QRNG & Speed & Interface  &Certifications \\
\hline \multirow{8}{*}{IDQ\cite{QuantisQ20:online}}
&Quantis-IDQ250C2 &250 Kbps&Chip& NIST SP800-22/90B, DieHarder\\
&Quantis-USB-4M &4 Mbps&USB&NIST SP800-22, CTL, METAS, AIS31\\
&Quantis-IDQ6MC1 &6 Mbps&Chip&NIST SP800-22/90A/B/C, DieHarder, AEC-Q100\\
&Quantis-IDQ20MC1 &20 Mbps&Chip&NIST SP800-22/90A/B/C, DieHarder\\
&Quantis-PCIe-16M&16 Mbps&PCLe&NIST SP800-22, CTL, METAS, BSI AIS 31\\
&Quantis-PCIe-40M &40 Mbps&PCLe&NIST SP800-22/90A/B/C, DieHarder\\
&Quantis-PCIe-240M &240 Mbps&PCLe&NIST SP800-22/90A/B/C, DieHarder\\
&Quantis-Appliance 2.0 &232 Mbps&Ethernet&NIST SP800-22/90B, DieHarder\\\hline
\multirow{1}{*}{PicoQuant \cite{QuantumR34:online}}
&PQRNG150&150 Mbps&USB&TESTU01\\\hline
\multirow{2}{*}{QuatumCTek \cite{QuantumR81:online}}
&QRNG100E&200 Mbps&USB&NIST SP800-22, GM/T 0005-2012\\
&QRNG100E&600 Mbps&Ethernet&NIST SP800-22, GM/T 0005-2012\\\hline
\multirow{3}{*}{ComScire \cite{Generato36:online} }
&PQ4000KS&4 Mbps& USB&ComScire QNGmeter\\
&PQ128MS&128 Mbps& USB&ComScire QNGmeter\\
&CS128M&128 Mbps& USB&ComScire QNGmeter\\\hline
\multirow{3}{*}{Quitessence Labs \cite{QuantumR74:online}}
&qStream 100&1 Gbps& Ethernet&NIST SP800-22/90A/B/C, DieHarder\\
&\multirow{2}{*}{qStream 200}&\multirow{2}{*}{1 Gbps}&\multirow{2}{*}{Ethernet}&NIST SP800-22/90A/B/C, DieHarder,\\ &&&&OASIS KMIP 1.0/1.1/1.2/1.3/1.4\\\hline
Quantum eMotion \cite{Products33:online} &QNG2&1 Gbps&Chip&NIST SP800-22, Diehard\\\hline
\multirow{3}{*}{EYL \cite{QuantumR31:online}}
&QRNG-H&1 Gbps& USB &NIST SP800-22/90B, BSI AIS 31\\
&QRNG-L&1 Mbps& USB &NIST SP800-22/90B, BSI AIS 31\\
&MQRNG&1 Gbps& PCLe &NIST SP800-22/90B, BSI AIS 31\\\hline
qutools \cite{quRNGdat53:online}&quRNG&50 Mbps& USB&NIST SP800-22, DieHarder\\\hline
\multirow{4}{*}{MPD \cite{MicroPho90:online}}
&QRN-16&16 Mbps&USB&NIST SP800-22, DieHarder, TESTU01\\
&QRN-32&32 Mbps&USB&NIST SP800-22, DieHarder, TESTU01\\
&QRN-64&64 Mbps&USB&NIST SP800-22, DieHarder, TESTU01\\
&QRN-128&128 Mbps&USB&NIST SP800-22, DieHarder, TESTU01\\\hline
\multirow{3}{*}{Quside \cite{quside:online}}
&Quside FMC 400&400 Mbps&Ethernet&Quside randomness metrology toolkit\\
&Quside PCLe 400&400 Mbps&PCLe&Quside randomness metrology toolkit\\
&Quside PCLe One&2 Gbps&PCLe&Quside randomness metrology toolkit\\\hline
QNU \cite{qnu}&TROPOS QNL-QRNG-X100&100 Mbps&Ethernet&NIST SP800-22, DieHard\\\hline
\end{tabular}}
\caption{Commercially available Quantum Random Number Generators.}
\label{tab:comtab}
\end{table}

This review has briefly compared the various methods for random number generation so that a user who wishes to employ random numbers in their schemes can make an informed decision. Further, it is difficult, if not impossible, to declare that a particular method is the best without knowing the necessities of the end-user. A large number of variables (e.g., bit generation rate, cost, ease of usage, availability, form factor, certifiability) need to be considered simultaneously for comparing any two or more schemes for QRNGs. In general, generators based on homodyne or heterodyne detection like the vacuum noise-based or ASE-based ones have significantly higher bit rates. Commercial products would be a better choice in terms of the form factor and ease of availability. Microchip or USB-based RNGs provide the ultimate portability, and most of such products come with accompanying software which makes it very easy to use for an end-user who may be a non-technical person. Device-independent methods have the highest certifiably albeit with lower generation rates. Along with the review of the generation methods for different classes of QRNGs, we have also tried to trace the origin of the randomness in the nonclassical sources of entropy, and that effort has provided a deeper insight into the world of randomness. This has provided completeness to this review, and it may provide some new insights to the researchers towards the ultimate goal of having a near-perfect random number generator.


\section*{Funding}
This work is  supported by grants received from DRDO, India via project number ANURAG/MMG/CARS/2018-19/071 and QUEST scheme of Interdisciplinary Cyber Physical Systems (ICPS) program of the Department of Science and Technology (DST), India (Grant No.: DST/ICPS/QuST/Theme-1/2019/14 (Q80)).

\section*{Acknowledgement}
Authors acknowledge Harsh Rathee and Kishore Thapliyal for their interest and feedback on this work.

\section*{Data availability}
No additional data is needed for this work.

\section*{Competing interests}
The authors declare that they have no competing interests.

\section*{Authors’ contributions}

All the authors have contributed equally. AP conceptualized the article and the foundational aspects of the problem. VM and SM performed a critical and rigorous analysis of the existing quantum random number generation schemes. All the authors contributed equally in the writing part to reveal the inherent symmetry leading to classification.

\bibliographystyle{vancouver} 
\bibliography{Random_number}      

\end{document}